\documentstyle[preprint,eqsecnum,aps]{revtex}
\begin{document}
\draft
\preprint{ROME1-1094/95, hep-ph/9503362}
\title{Neutralino Production as {\sc SuSy} Discovery \protect\\
Process at CERN LEP2}
\author{Sandro Ambrosanio\thanks{
{\it e-mail:} {\tt ambrosanio@roma1.infn.it}}
and Barbara Mele\thanks{
{\it e-mail:} {\tt mele@roma1.infn.it}}}
\address{Dipartimento di Fisica, Universit\`a ``La Sapienza''
and I.N.F.N., Sezione di Roma, \protect\\
P.le Aldo Moro 2, I-00185 Rome, Italy}
\date{June 5, 1995}
\maketitle
\begin{abstract}
A thorough study is performed on pair production and
signatures of supersymmetric neutralinos in the MSSM at
LEP2. Particular attention is paid to the region of the {\sc
SuSy} parameter space where the associated production of
lightest and next-to-lightest neutralinos is the only visible
allowed supersymmetric process. In that region, the signal is
critically dependent on the selectron masses
$m_{\tilde{e}_{L,R}}$. For $\sqrt{s}/2 < m_{\tilde{e}_{L,R}}
\;\raisebox{-.5ex}{\rlap{$\sim$}}
\raisebox{.5ex}{$<$}\;200$-300 GeV and charginos above the
threshold for pair production, neutralinos arising from
$e^{\scriptscriptstyle +}e^{\scriptscriptstyle -}
\rightarrow \tilde{\chi}^{\scriptscriptstyle 0}_{1}
\tilde{\chi}^{\scriptscriptstyle 0}_{2}$ could be the only
{\sc SuSy} signal detectable at LEP2.
\end{abstract}

\newpage
\thispagestyle{empty}
\null
\newpage

\setcounter{page}{1}

\section{Introduction}
\label{sec:intro}

If supersymmetry ({\sc SuSy}) is introduced to solve the
naturalness problem encountered when embedding the Standard
Model (SM) in a Grand Unified Theory (GUT), one is forced to
assume that super-partner masses are not much larger than
the scale of the electroweak (EW) symmetry breaking. In
particular, the s-partners of the particles that interact
only electroweakly should be in the lower range of the {\sc SuSy}
mass spectrum. These particles are more efficiently produced
at $ e^{ \scriptscriptstyle  +}e^{ \scriptscriptstyle  -} $
colliders where there is no large QCD background. The
lightest s-partners of the EW gauge and Higgs bosons will be
most probably the first to be accessible in
$ e^{\scriptscriptstyle  +}e^{ \scriptscriptstyle  -} $
collisions (see,  \hbox{e.g.}{},
\hbox{Ref.~}{}\cite{kane93}).

We restrict ourselves to the framework of the Minimal
Supersymmetric Standard Model (MSSM) with universality
conditions on soft {\sc SuSy}-breaking parameters at the GUT
scale and $R$-parity unbroken \cite{reviews,hk}. In the most
likely scenarios, the lightest {\sc SuSy} particle, that in the
hypothesis of conserved $R$-parity is stable, is the
lightest neutralino. In this model, all masses and couplings
are set by choosing the values of a finite set of parameters
at the GUT scale, usually $m_0$ (the common scalar mass),
$m_{1/2}$ (the common gaugino mass), $\mu$ (the {\sc SuSy}
Higgs-mixing mass), $ \tan \beta $ (the ratio of vacuum
expectation values for the two Higgs doublets) and $A_0$
(the common soft-breaking scalar trilinear coupling). Two
further parameters (\hbox{e.g.}{}\ $ m_{A^{
\scriptscriptstyle  0}} $, $B_0$) are needed to describe the
Higgs sector if one does not use relations coming from the
requirement that the radiative EW-symmetry breaking take
place at the correct scale.

At LEP2, one could produce sfermion pairs and/or
chargino/neutralino pairs. Charged sfermions and charginos,
when allowed by phase-space, are the easiest s-particles to
produce since they are always directly coupled to photon and
$ Z^{ \scriptscriptstyle  0} $ vector bosons. On the other
hand, in general the lightest neutralino states are lighter
than charginos and sfermions, but they are linear
combinations of neutral gauginos and higgsinos (hence not
coupled to photons) that can decouple also from the
$ Z^{\scriptscriptstyle  0} $ and consequently have lower
production cross sections.

In this paper, we concentrate on neutralino production at
LEP2. We consider with particular attention the regions of
{\sc SuSy}-parameter space where sfermions and charginos are above
the pair-production threshold at LEP2 (\hbox{i.e.}{}, they
have masses larger than about $ M_{ \scriptscriptstyle  Z}
$), while the Lightest (LN) and Next-to-Lightest (NLN)
Neutralinos ($ \tilde{\chi}^{ \scriptscriptstyle  0}_1 $ and
$ \tilde{\chi}^{ \scriptscriptstyle  0}_2 $, respectively)
can be produced through the process:
\begin{equation}
e^{ \scriptscriptstyle  +}e^{ \scriptscriptstyle  -}
\rightarrow   \tilde{\chi}^{ \scriptscriptstyle  0}_1
\tilde{\chi}^{ \scriptscriptstyle  0}_2. \label{eq:proc}
\end{equation}
A spectacular signature is associated to this
channel, where $ \tilde{\chi}^{ \scriptscriptstyle  0}_1 $
goes out of the experimental apparatus undetected and the
jets of particles coming from the
$ \tilde{\chi}^{\scriptscriptstyle  0}_2 $ decay is mostly
unbalanced in energy and momentum. We do not consider production
of lightest neutralino pairs ($ e^{ \scriptscriptstyle  +}
e^{\scriptscriptstyle  -}    \rightarrow
\tilde{\chi}^{\scriptscriptstyle  0}_1
\tilde{\chi}^{ \scriptscriptstyle 0}_1$) since it gives rise
to an invisible signal at Born level.
The process (\ref{eq:proc}) has been carefully
studied at LEP1 energies, where the absence of a neutralino
signal extends the regions of {\sc SuSy} parameters excluded by
direct search and contributions to the $ Z^{
\scriptscriptstyle  0} $ width of chargino production
\cite{lep1neut}. Analogously, we want to study the potential
of process (\ref{eq:proc}) at LEP2 to probe regions of the
parameter space not covered by chargino searches (we will
name these regions Neutralino Regions (NR)). To this aim, we
carry out an exhaustive analysis of cross sections and decay
rates corresponding to all possible signatures in the MSSM,
up-dating and complementing previous partial studies
\cite{bartl86a,dionisi}. Some results relative to heavier
neutralino pair production will be also presented, when
relevant in the Neutralino Regions.

The reaction (\ref{eq:proc}) proceeds through two mechanisms
(\hbox{cfr.}{}\  \hbox{Fig.~}{}\ref{fig:feyprod}): an
$s$-channel $ Z^{ \scriptscriptstyle  0} $ exchange and a
$t(u)$-channel (either left or right) selectron exchange.
Only higgsino components of neutralinos (that directly
couple to $ Z^{ \scriptscriptstyle  0} $) have a r\^ole in
the $s$-channel. On the contrary, in the limit of negligible
electron mass, only photino and Z-ino components take parts
into the $t$-channel diagrams.

At LEP1, the relevant $ \tilde{\chi}^{ \scriptscriptstyle
0}_i $ components are the higgsino ones, due to the
$ Z^{\scriptscriptstyle  0} $ resonance, and cross sections
are fixed by only three parameters: $M_2$, $\mu$ and
$\tan \beta $. On the contrary, at LEP2 the $s$-channel cross
sections in the Neutralino Regions can be smaller than
$t$-channel ones and hence the selectron mass becomes a
crucial parameter too.

Besides considering the continuous parameter dependence of
our results, some particularly meaningful scenarios will be
identified and studied in the Neutralino-Region parameter
space.

In Section \ref{sec:matrix}, we set notations by recalling
the neutralino and chargino mass matrices. We also study the
physical components and mass spectrum of the two lightest
neutralinos as functions of MSSM parameters. Furthermore, we
define the Neutralino Regions and describe their interest.
In Section \ref{sec:prod}, we study
$\tilde{\chi}^{ \scriptscriptstyle  0}_1
\tilde{\chi}^{ \scriptscriptstyle  0}_2 $-production cross
sections at LEP2 and identify a set of significant scenarios
for a systematic study. In Section \ref{sec:decay}, some results
concerning the main $ \tilde{\chi}^{ \scriptscriptstyle 0}_2$
neutralino decays are reported. A more in-depth
investigation on all possible
$ \tilde{\chi}^{\scriptscriptstyle  0}_2 $ decay channels can
be found in \hbox{Ref.~}{}\cite{ambmele}. Finally, in Section
\ref{sec:totrates},
total rates for all relevant signatures at LEP2 coming from
$\tilde{\chi}^{ \scriptscriptstyle  0}_1
\tilde{\chi}^{\scriptscriptstyle  0}_2 $ production are studied
in detail.
In Appendix \ref{app:a}, we give formulae that relate scalar
masses to the MSSM parameters through renormalization-group
equations (RGE's).

\section{The electroweak gaugino/higgsino sector}
\label{sec:matrix}

In the MSSM, four fermionic partners of the neutral
components of the SM gauge and Higgs bosons are predicted:
the photino $ \tilde{\gamma} $, the Z-ino $ \tilde{Z} $
(mixtures of the U(1) $\tilde{B}$ and SU(2) $\tilde{W}_3$
gauginos), and the two higgsinos $ \tilde{H}^{
\scriptscriptstyle  0}_1 $ and $ \tilde{H}^{
\scriptscriptstyle  0}_2 $ (partners of the two
Higgs-doublet neutral components). In general, this
interaction eigenstates mix, their mixing being controlled
by a mass matrix $Y$ \cite{neumatr,frekan} defined by:
\begin{mathletters}
\label{eq:neumixing}
\begin{equation}
{\cal L} ^0_M = -\frac{1}{2} \psi^0_i Y_{ij} \psi^0_j +
h.c., \label{eq:neumasslag}
\end{equation}
where:
\begin{equation}
Y = \left( \begin{array}{cccc} M_2
\sin^2\!\theta_{ \scriptscriptstyle  W}  + M_1
\cos^2\!\theta_{ \scriptscriptstyle  W}  & (M_2 - M_1)
\sin\theta_{ \scriptscriptstyle  W}  \cos\theta_{
\scriptscriptstyle  W}       & 0    & 0        \\ (M_2 -
M_1) \sin\theta_{ \scriptscriptstyle  W}  \cos\theta_{
\scriptscriptstyle  W}     & M_2  \cos^2\!\theta_{
\scriptscriptstyle  W}  + M_1  \sin^2\!\theta_{
\scriptscriptstyle  W}     &  M_{ \scriptscriptstyle  Z}   &
0        \\     0      &       M_{ \scriptscriptstyle  Z} &
       \mu  \sin 2\beta           &  -\mu  \cos 2\beta
\\     0      &      0       &        -\mu  \cos 2\beta
&  -\mu  \sin 2\beta     \\ \end{array} \right).
\label{eq:neumat}
\end{equation}
\end{mathletters}
Closely following the notations of
\hbox{Refs.~}{}\cite{bartl86a,bartl89},
\hbox{Eqs.~}{}(\ref{eq:neumixing}) are written by suitably
choosing the basis:
\begin{equation}
\psi^0_j =
(-i\phi_{\gamma}, -i\phi_Z, \psi^a_H, \psi^b_H), \; \; \; \;
j = 1, \ldots 4, \label{eq:base}
\end{equation}
where:
\begin{eqnarray*}
\psi^a_H & = & \psi^1_{H_1}  \cos \beta  -
\psi^2_{H_2}  \sin \beta  \; , \\ \psi^b_H & = &
\psi^1_{H_1}  \sin \beta  + \psi^2_{H_2}  \cos \beta  \; ,
\end{eqnarray*}
and $\phi_{\gamma}$, $\phi_Z$,
$\psi^1_{H_1}$, $\psi^2_{H_2}$ are two-component
spinorial-fields. In  \hbox{Eq.~}{}(\ref{eq:neumat}), $ \tan
\beta  = \frac{v_{ \scriptscriptstyle  2}}{v_{
\scriptscriptstyle 1}}$ and $M_{1,2}$ are the U(1)- and
SU(2)-gaugino masses at the EW scale. By assuming
gaugino-mass unification at $M_{\rm  \scriptscriptstyle
GUT}$, $M_1$ can be related to $M_2$ by the equation:
\begin{equation}
M_1 = \frac{5}{3} \tan^2\theta_{
\scriptscriptstyle  W} M_2, \label{eq:unigau}
\end{equation}
that arises from one-loop RGE's (\hbox{cfr.}{}\ Appendix
\ref{app:a}). The $Y$ matrix (that, excluding $CP$ violations
in this sector of the model, is real and symmetric) can be
diagonalized by a unitary $4 \times 4$ matrix $N$:
\[ N_{im}N_{kn} Y_{mn} =  m_{\tilde{\chi}^0_i}  \delta_{ik}, \]
where
$ m_{\tilde{\chi}^0_i} $ \ ($i=1, \ldots 4$) is the mass
eigenvalue relative to the $i$-th neutralino state, given by
the two-component spinor field $\chi^{ \scriptscriptstyle
0}_i = N_{ij} \psi^{ \scriptscriptstyle  0}_j$. Then,
\hbox{Eq.~}{}(\ref{eq:neumasslag}) can be rewritten, by
using the four-component neutral Majorana-spinor formalism,
in the form:
\[  {\cal L} ^0_M = -\frac{1}{2} \sum_i
m_{\tilde{\chi}^0_i}  \bar{ \tilde{\chi}^{
\scriptscriptstyle  0}_i }  \tilde{\chi}^{
\scriptscriptstyle  0}_i , \]
where:
\[  \tilde{\chi}^{\scriptscriptstyle  0}_i
= \left( \begin{array}{c}    \chi^0_i          \\
\bar{\chi}^0_i    \end{array}
\right). \]
The $N$ matrix can be chosen real and
orthogonal. In this case some of the $ m_{\tilde{\chi}^0_i}$
eigenvalues can be negative. The sign of
$m_{\tilde{\chi}^0_i} $ is related to the $CP$ quantum number
of the $i$-th neutralino \cite{hk,bartl86a,petcov}. By
solving a 4-th degree eigenvalue equation, one can find the
expressions of $ m_{\tilde{\chi}^0_i} $ and of physical
composition of the corresponding eigenstate in terms of the
independent parameter set $\mu$, $M_2$ and $ \tan \beta $ (a
complete treatment can be found in
\hbox{Ref.~}{}\cite{bartl89}).

In our analysis, we study the small and moderate range of $
\tan \beta $. This may be also interesting in connection
with scenarios with the top mass at its infrared fixed point
\cite{ref:fixpoint}. In particular, we set the value of $
\tan \beta $ at either 1.5 or 4.

Here, we are concerned mainly with the two lightest
neutralino states ($i =1,2$). In
\hbox{Figs.~}{}\ref{mix:n1gau}--\ref{mix:n2hino}, the
behaviour of their gaugino and higgsino components
(\hbox{i.e.}{}, the square moduli of $N_{ij}$, for $i=1,2$
and $j=1, \ldots 4$) is shown for $ \tan \beta  = 1.5$
versus $\mu$ and $M_2$.

By looking at the $Y$ matrix in
\hbox{Eq.~}{}(\ref{eq:neumat}), one can easily realize that
for either $|\mu| \gg  M_{ \scriptscriptstyle  Z} $ or $M_2
\gg  M_{ \scriptscriptstyle  Z} $ the $2 \times 2$ blocks
relative to the gaugino and higgsino sectors do not mix and
the two lightest neutralinos get either both gauginos (with
masses close to $M_1$ and $M_2$) or both higgsinos (with
degenerate masses close to $\pm |\mu|$).

In  \hbox{Fig.~}{}\ref{mix:n1gau}, one can note that the
photino component of the lightest neutralino is dominant for
$M_2  \;\raisebox{-.5ex}{\rlap{$\sim$}}
\raisebox{.5ex}{$<$}\;   M_{ \scriptscriptstyle  Z} /2$ and
$\mu \neq 0$, and in regions where $M_2
\;\raisebox{-.5ex}{\rlap{$\sim$}} \raisebox{.5ex}{$<$}\;
-2\mu$ with $\mu < 0$, while the $ \tilde{Z} $ one is
particularly enhanced in the positive-$\mu$ half-plane for
low $M_2$.  \hbox{Fig.~}{}\ref{mix:n1hino} shows that
higgsino components of $ \tilde{\chi}^{ \scriptscriptstyle
0}_1 $ are very important in the $2|\mu|
\;\raisebox{-.5ex}{\rlap{$\sim$}} \raisebox{.5ex}{$<$}\;
M_2$ triangle region. In this area, the $ \tilde{H}^{
\scriptscriptstyle  0}_a $ component dominates for $\mu > 0$
and $M_2  \;\raisebox{-.5ex}{\rlap{$\sim$}}
\raisebox{.5ex}{$>$}\;   M_{ \scriptscriptstyle  Z} $,
whereas the lightest neutralino is nearly a pure $
\tilde{H}^{ \scriptscriptstyle  0}_b $ for $\mu \le 0$.

The physical composition of
$ \tilde{\chi}^{\scriptscriptstyle  0}_2 $ can be
observed in \hbox{Figs.~}{}\ref{mix:n2gau} (gauginos)
and \ref{mix:n2hino} (higgsinos). The photino component in
$\tilde{\chi}^{ \scriptscriptstyle  0}_2 $ is sizeable only
for $2\mu  \;\raisebox{-.5ex}{\rlap{$\sim$}}
\raisebox{.5ex}{$>$}\;  M_2$, if $M_2
\;\raisebox{-.5ex}{\rlap{$\sim$}} \raisebox{.5ex}{$>$}\;
M_{ \scriptscriptstyle  Z} /2$, while a large $ \tilde{Z} $
component can be found in the negative-$\mu$ half-plane for
$|\mu| \gg M_2$ or, to a lesser extent, in the
positive-$\mu$ half-plane for $M_2$ between
$ M_{\scriptscriptstyle  Z} $ and
$2 M_{ \scriptscriptstyle  Z}$, provided $\mu$ is
positive and large enough.
\hbox{Fig.~}{}\ref{mix:n2hino} shows that the higgsino
composition of the next-to-lightest neutralino is somehow
complementary to the lightest neutralino one. The main
features of the above pictures keep valid when increasing
$\tan \beta $ up to 4.

Concerning the mass spectrum of light neutralinos, contour
plots for $| m_{\tilde{\chi}^0_i} |$ in GeV are given in
\hbox{Fig.~}{}\ref{neut12massign} for
$ \tilde{\chi}^{ \scriptscriptstyle  0}_1 $ and
$ \tilde{\chi}^{\scriptscriptstyle  0}_2 $, with
$ \tan \beta  = 1.5$. Dark area represents regions in the
$(\mu, M_2)$ plane where the mass eigenvalue of the neutralino
is negative.

The $ m_{\tilde{\chi}^0_1}  < 0$ area in
\hbox{Fig.~}{}\ref{neut12massign} is bounded by the
$\mu = 0$ line (along which
$ \tilde{\chi}^{ \scriptscriptstyle 0}_1 $ is identical to
$ \tilde{H}^{ \scriptscriptstyle 0}_b $ and massless), by
the hyperbole
$\mu M_1 M_2 = (M_2 \sin^2\!\theta_{ \scriptscriptstyle  W}
+ M_1 \cos^2\!\theta_{ \scriptscriptstyle  W} )
M_{\scriptscriptstyle  Z} ^2  \sin 2\beta $ \  \hbox{i.e.}{},
by using  \hbox{Eq.~}{}(\ref{eq:unigau}), $\mu M_2 =
\frac{8}{5}  M_{ \scriptscriptstyle  W} ^2  \sin 2\beta $
(along which the $ \tilde{\chi}^{ \scriptscriptstyle  0}_1 $
is a massless non-trivial mixed state, mostly made of $
\tilde{Z} $ and $\tilde{H}$) and, in the lowest part of the
positive-$\mu$ half-plane, by the contour for crossing of
the two lightest neutralino masses, that is complementary to
the one for $ m_{\tilde{\chi}^0_2} $. A similar effect
occurs for the $ \tilde{\chi}^{ \scriptscriptstyle  0}_2 $-$
\tilde{\chi}^{ \scriptscriptstyle  0}_3 $ crossing, as the
dark ``V'' region in the $ m_{\tilde{\chi}^0_2} $ contour
plot shows. In that region, approximatively bounded by the
$M_2 = \pm \mu +  M_{ \scriptscriptstyle  Z} $ lines, the
higgsino components dominate both $ \tilde{\chi}^{
\scriptscriptstyle  0}_1 $ and $ \tilde{\chi}^{
\scriptscriptstyle  0}_2 $. Again, when increasing $ \tan
\beta $ up to 4, one does not observe substantial
variations.

In the following, it will be useful to consider also the
chargino sector. The corresponding mass term in the
Lagrangian is \cite{hk,bartl92}:
\begin{mathletters}
\label{lag:chamass}
\begin{eqnarray} {\cal L}^{\pm}_M & = &
- \frac{1}{2} (\psi^+ \; \psi^-) \left( \begin{array}{cc}
	0 & X^T \\    			X & 0   \\
\end{array} \right) \left( \begin{array}{c} 	\psi^+ \\
		  		\psi^- \\ \end{array}
\right) + h.c. \; , \label{lagchamass} \\ X 		 & =
&  \left( \begin{array}{cc} M_2 	      &  M_{
\scriptscriptstyle  W}   \sqrt{2}   \sin \beta  \\  M_{
\scriptscriptstyle  W}   \sqrt{2}   \cos \beta  & \mu
    \end{array} \right), \label{eq:chamat}
\end{eqnarray}
\end{mathletters}
where $\psi^+_j = (-i\phi^+, \;
\psi^1_{H_2})$, $\psi^-_j = (-i \phi^-, \; \psi^2_{H_1})$, \
$j = 1,2$ and $\phi^{\pm}$, $\psi^1_{H_2}$, $\psi^2_{H_1}$
are two-component spinorial-fields of W-inos and charged
higgsinos, respectively. The mass matrix $X$ can be
diagonalized by two $2 \times 2$ unitary matrices $U$ and
$V$:
\[ U_{im} V_{jn} X_{mn} = m_{\tilde{\chi}^{\pm}_i} \delta_{ij}, \]
where $m_{\tilde{\chi}^{\pm}_i}$ is the mass eigenvalue for the
$i$-th chargino state, which is defined by:
$\chi^+_i = V_{ij} \psi^+_j$, $\chi^-_i = U_{ij} \psi^-_j$,
\ $i,j = 1,2$ \ ($V$ and $U$ are taken real after assuming $CP$
conservation). Here $\chi^{+(-)}_i$ are the two-component
spinors corresponding to the positive- (negative-) charged
part of the four-component Dirac-spinor of
$ \tilde{\chi}^{ \scriptscriptstyle  \pm}_i $. After diagonalization,
one is able to derive a simple formula for the chargino-mass
eigenvalues:
\begin{equation}
m_{\tilde{\chi}^{\pm}_{1,2}} =
\frac{1}{2}
\left[\sqrt{(M_2 -\mu)^2+2 M_{ \scriptscriptstyle  W} ^2(1+
\sin 2\beta )} \mp \sqrt{(M_2+\mu)^2+2 M_{
\scriptscriptstyle  W} ^2(1- \sin 2\beta )} \right]
\label{charmass}
\end{equation}

At this point, it is straightforward to set regions in the
$(\mu, M_2)$ plane where $ \tilde{\chi}^{ \scriptscriptstyle
 0}_1  \tilde{\chi}^{ \scriptscriptstyle  0}_2 $ production
is allowed by phase-space, but chargino-pair production is
not, for which one has:
\begin{equation}
m_{\tilde{\chi}^0_1}  +  m_{\tilde{\chi}^0_2}  <  \sqrt{s}
< 2  m_{\tilde{\chi}^{\pm}_1} .
\label{nrcond}
\end{equation}

These regions are shown in
\hbox{Figs.~}{}\ref{fig:scentgb15}, \ref{fig:scentgb4} for $
\tan \beta =1.5$, 4 and $ \sqrt{s} =190$ GeV
(curves `N190' correspond to $ \sqrt{s} =
m_{\tilde{\chi}^0_1} + m_{\tilde{\chi}^0_2} $, while curves
`C190' are for $ \sqrt{s} = 2  m_{\tilde{\chi}^{\pm}_1} $).
We will call NR$^{\pm}$ the two disconnected regions where
\hbox{Eq.~}{}(\ref{nrcond}) holds and $\mu
\;\raisebox{-.5ex}{\rlap{$<$}} \raisebox{.5ex}{$>$}\;  0$.
We can see that there is a conspicuous increase in the
accessible parameter space due to the lower neutralino $
\tilde{\chi}^{ \scriptscriptstyle  0}_1  +  \tilde{\chi}^{
\scriptscriptstyle  0}_2 $ threshold with respect to
chargino pairs. The relevant portions of space are placed,
for our choice of $ \tan \beta $ values, where $\mu
\;\raisebox{-.5ex}{\rlap{$\sim$}} \raisebox{.5ex}{$<$}\;  -
M_{ \scriptscriptstyle  Z} $ and $\mu
\;\raisebox{-.5ex}{\rlap{$\sim$}} \raisebox{.5ex}{$>$}\;
1.5  M_{ \scriptscriptstyle  Z} $. For $ \tan \beta  = 1.5$
(\hbox{Fig.~}{}\ref{fig:scentgb15}), the NR$^-$ is
centered around $M_2 = 1.1  M_{ \scriptscriptstyle  Z} $,
while the NR$^+$ is slightly shifted to higher $M_2$
values. By increasing $ \tan \beta $ (
\hbox{Fig.~}{}\ref{fig:scentgb4}), the asymmetry in the two
regions decreases. The shaded area shows the region excluded
by LEP1.

We do not consider in this work the small modifications of
the above general scenario that could arise from radiative
corrections to gaugino/higgsino masses. Recent calculations
\cite{lahanas93,pierce93} at the one-loop level give
indication for typical corrections of the order of 6\% (or
somewhat higher in particular cases for the lightest
neutralino) with same sign for all neutralino/chargino
states. So, they do not change the relative configuration of
neutralino and chargino masses and do not affect our general
discussion. Also, such corrections are of the same order of
magnitude as other neglected effects,  \hbox{e.g.}{}\ other
threshold effects in the RGE evolution.

\section{Neutralino cross sections at LEP2}
\label{sec:prod}

In this section, we study total cross section for the
process $ e^{ \scriptscriptstyle  +}e^{ \scriptscriptstyle
-}    \rightarrow   \tilde{\chi}^{ \scriptscriptstyle  0}_1
\tilde{\chi}^{ \scriptscriptstyle  0}_2$ at LEP2. The
relevant formulae needed to compute neutralino total cross
sections can be found in  \hbox{Ref.~}{}\cite{bartl86a}.
Crucial parameters in the prediction of total rates are the
values of selectron masses $ m_{\tilde{e}_{L,R}} $ that
enter the $t$-channel amplitudes. These are directly related
to $m_0$ through the RGE's that govern the running of scalar
masses from the GUT scale down to $ M_{ \scriptscriptstyle
Z} $ (see Appendix \ref{app:a}). Then, one can compute rates
for different signals coming from the
$ \tilde{\chi}^{\scriptscriptstyle  0}_2 $ decay as
functions of $m_0$, since, once $m_0$ is fixed, all other
scalar particles entering the
$\tilde{\chi}^{\scriptscriptstyle 0}_2$ decays (excluding Higgses)
are set too, for any $M_2$ and $\tan \beta $ value.

The formulae that connect all relevant scalar masses to
$m_0$ are collected in Appendix \ref{app:a}, where more details
about approximations and strategies for evaluating the sfermion
spectrum can also be found. Since we are particularly
interested in studying regions of the parameter space in
which no pair-production processes of {\sc SuSy} particles are
allowed other than neutralino production, we choose to
perform most of our analysis in scenarios with $m_0 \ge
M_{\scriptscriptstyle  Z} $. This choice has two important
consequences. Firstly, for $M_2$ not too small
(\hbox{e.g.}{}, $M_2$ values not excluded by LEP1 data), it
gives rise to scalar masses greater than the LEP2 beam
energy, \hbox{i.e.}{}\ scalars can not be pair-produced at
LEP2 (\hbox{cfr.}{}\ Appendix \ref{app:a}). Secondly, with
these relatively heavy scalars, the two-body decays
$\tilde{\chi}^{ \scriptscriptstyle  0}_2   \rightarrow
f\tilde{f}_{L,R}$ are in most cases not allowed. This point
will be resumed in Sections \ref{sec:decay} and \ref{sec:totrates}.

A general feature of
$ \tilde{\chi}^{ \scriptscriptstyle 0}_1
\tilde{\chi}^{ \scriptscriptstyle  0}_2 $ cross
sections is that, in order to have a large contribution
either from the $s$-channel or the $t$-channel
(\hbox{cfr.}{}\  \hbox{Fig.~}{}\ref{fig:feyprod}), both $
\tilde{\chi}^{ \scriptscriptstyle  0}_1 $ and
$\tilde{\chi}^{ \scriptscriptstyle  0}_2 $ should have a
large component of either higgsino or gaugino. Nevertheless,
the $t$-channel contribution will be in general lower,
especially when the selectron masses in the $t$-channel
propagators are assumed larger than $ M_{ \scriptscriptstyle
 Z} $. Mixed cases, where the two neutralinos have different
dominant components, give rise in general to comparable
contribution from $s$, $t$ amplitudes and their relative
interference. The limit of production of one pure higgsino
plus one pure gaugino is dynamically forbidden and has null
cross section (for $m_e = 0$). These different cases will be
discussed in what follows.

In  \hbox{Fig.~}{}\ref{n1n2prodm91tgb15rs190}, the contour
plot of the total cross section (in fb) for
$ e^{\scriptscriptstyle  +}e^{ \scriptscriptstyle  -}
\rightarrow   \tilde{\chi}^{ \scriptscriptstyle  0}_1
\tilde{\chi}^{ \scriptscriptstyle 0}_2$ is shown for $ \tan
\beta  = 1.5$ , $m_0 =  M_{ \scriptscriptstyle  Z} $ (that
\hbox{e.g.}{}, for $M_2 =  M_{ \scriptscriptstyle  Z} $,
corresponds to $ m_{\tilde{e}_L} = 124$ GeV and $
m_{\tilde{e}_R}  = 104$ GeV,  \hbox{cfr.}{}\
\hbox{Table~}{}\ref{tab:scentgb1p5}) and $ \sqrt{s}  = 190$
GeV. We can distinguish different regions in the
$(\mu, M_2)$ plane on the basis of the magnitude of the
total cross section. The largest rates (up to about 2 pb)
are reached for $|\mu|  \;\raisebox{-.5ex}{\rlap{$\sim$}}
\raisebox{.5ex}{$<$}\;   M_{ \scriptscriptstyle  Z} $ and
$M_2  \;\raisebox{-.5ex}{\rlap{$\sim$}}
\raisebox{.5ex}{$>$}\;   M_{ \scriptscriptstyle  Z} $, where
the two lightest neutralinos are mainly higgsinos (with
masses close to $\pm |\mu|$) and hence are fully coupled to
the $ Z^{ \scriptscriptstyle  0} $ in the $s$-channel. In
what follows, we will name the two regions in this area, on
the left and on the right of the LEP1 excluded region,
HCS$^-$ and HCS$^+$ (standing for High Cross Section)
regions, respectively. These regions are shown in
\hbox{Fig.~}{}\ref{fig:scentgb15}, where the contour plot
for
$\sigma( e^{ \scriptscriptstyle  +}e^{\scriptscriptstyle  -}
\rightarrow \tilde{\chi}^{\scriptscriptstyle  0}_1
\tilde{\chi}^{ \scriptscriptstyle 0}_2) = 1$ pb
is also plotted, for $m_0 =  M_{\scriptscriptstyle  Z} $.

In the regions NR$^+$ and NR$^-$, the gaugino components
and the related $t$-channel contribution to the total cross
section come into play and cross sections drop. Typical
total rates in the regions NR$^-$ are of the order of
$50\div 100$ fb, corresponding to a number of about
$25 \div 50$ events, for an integrated luminosity of 500
pb$^{-1}$. Somewhat lower cross sections are observed
in the NR$^+$ region, where, due to the higher value of
$M_2$, heavier selectrons are exchanged in the $t$-channel
(\hbox{cfr.}{}\  \hbox{Eqs.~}{}(\ref{eq:msf}),
(\ref{eq:sfmass}) and  \hbox{Table~}{}\ref{tab:scentgb1p5}).

In  \hbox{Fig.~}{}\ref{n1n2prodm273tgb15rs190}, we show the
effect of rising $m_0$ up to $3 M_{ \scriptscriptstyle  Z}
$. With respect to
\hbox{Fig.~}{}\ref{n1n2prodm91tgb15rs190}, cross sections
are considerably reduced in regions where $t$-channel
amplitudes are relevant. For instance, in the Neutralino
Regions
$\sigma( e^{ \scriptscriptstyle  +}e^{\scriptscriptstyle  -}
\rightarrow \tilde{\chi}^{\scriptscriptstyle  0}_1
\tilde{\chi}^{ \scriptscriptstyle 0}_2)$ is at most of the
order of $20\div 30$ fb.
A moderate change is observed when varying the value of
$ \tan \beta $ up to 4 (\hbox{Fig.~}{}\ref{n1n2prodm91tgb4rs190}),
due mainly to the different NR$^+$ and NR$^-$ position and
shape in the $(\mu, M_2)$ plane.

In order to clarify the origin of the total cross-section
behaviour in the $(\mu,M_2)$ plane and, in particular, in
the NR$^{\pm}$ and HCS$^{\pm}$ regions, we consider now in
detail a set of specific cases in the parameter space. In
\hbox{Table~}{}\ref{tab:scentgb1p5}, we report the following
features for six different scenarios (A, B, C, D in the
Neutralino Regions and H$^{\pm}$ in the High Cross Section
regions), defined by their values of $\mu$ and $M_2$ for $
\tan \beta =1.5$: \\
i) values of neutralino and chargino
masses (including the correct sign); \\
ii) the percentage components of different gaugino and
higgsino physical states for \\
$ \tilde{\chi}^{ \scriptscriptstyle  0}_1 $ and
$ \tilde{\chi}^{ \scriptscriptstyle  0}_2 $; \\
iii) scalar masses arising from $m_0=M_{\scriptscriptstyle Z}$
and RGE-evolution, calculated by  \\
\hbox{Eqs.~}{}(\ref{eq:msf}) and (\ref{eq:sfmass}); \\
iv) total cross sections (in fb), for $m_0 =
M_{\scriptscriptstyle Z}$ and $\sqrt{s} = 190$ GeV, of
all the allowed neutralino-pair production processes:
$e^{ \scriptscriptstyle  +}e^{\scriptscriptstyle  -}  \rightarrow
\tilde{\chi}^{\scriptscriptstyle  0}_1
\tilde{\chi}^{ \scriptscriptstyle 0}_1 , \
\tilde{\chi}^{ \scriptscriptstyle  0}_1
\tilde{\chi}^{ \scriptscriptstyle  0}_2 $ and, when below
threshold, $ \tilde{\chi}^{ \scriptscriptstyle  0}_1
\tilde{\chi}^{ \scriptscriptstyle  0}_3 , \
\tilde{\chi}^{\scriptscriptstyle  0}_1
\tilde{\chi}^{ \scriptscriptstyle 0}_4 , \
\tilde{\chi}^{ \scriptscriptstyle  0}_2
\tilde{\chi}^{ \scriptscriptstyle  0}_2 $; \\
v) for the main
$ \tilde{\chi}^{ \scriptscriptstyle  0}_1
\tilde{\chi}^{ \scriptscriptstyle  0}_2 $ channel, different
contributions to the total rates coming from $s$-channel,
$(t+u)$-channels and $(st+su)$ interferences. \\
The location of the six points in the $(\mu, M_2)$ plane is
shown in  \hbox{Fig.~}{}\ref{fig:scentgb15}. \\
We now analyse the physical features of these scenarios. In
Scenario A, where $\mu = -3  M_{ \scriptscriptstyle  Z} $
and $M_2 =  M_{ \scriptscriptstyle  Z} $, the lightest
neutralino is mostly a gaugino with a predominance of
photino. The next-to-lightest neutralino is still a gaugino,
but with inverse $ \tilde{\gamma} - \tilde{Z} $ relative
composition. In this case, since $|\mu| \gg M_2,  M_{
\scriptscriptstyle  Z} $, masses obey the asymptotic
relation $ m_{\tilde{\chi}^{\pm}_1}  \simeq
m_{\tilde{\chi}^0_2} \simeq 2  m_{\tilde{\chi}^0_1} $. In
such a scenario, the $ \tilde{\chi}^{ \scriptscriptstyle
0}_1  \tilde{\chi}^{ \scriptscriptstyle  0}_2 $ cross
section comes uniquely from $t$- and $u$-channels and is
about 146 fb. In Scenario B, where $\mu = - M_{
\scriptscriptstyle  Z} $ and $M_2 =  M_{ \scriptscriptstyle
Z} $, there is a mixed situation, where $ \tilde{\chi}^{
\scriptscriptstyle  0}_1 $ is predominantly a photino, while
$ \tilde{\chi}^{ \scriptscriptstyle  0}_2 $ is mostly an $
\tilde{H}^{ \scriptscriptstyle  0}_b $. The total $
\tilde{\chi}^{ \scriptscriptstyle  0}_1  \tilde{\chi}^{
\scriptscriptstyle  0}_2 $ production rate receives
contributions from all channels and interferences and is
about 113 fb. One can notice that in this case the
production cross section for $ \tilde{\chi}^{
\scriptscriptstyle  0}_1  \tilde{\chi}^{ \scriptscriptstyle
0}_3 $ pairs is larger than for $ \tilde{\chi}^{
\scriptscriptstyle  0}_1  \tilde{\chi}^{ \scriptscriptstyle
0}_2 $, although $ m_{\tilde{\chi}^0_3} $ is quite heavier
than $ m_{\tilde{\chi}^0_2} $. This is due to the different
composition of $ \tilde{\chi}^{ \scriptscriptstyle  0}_3 $,
\hbox{i.e.}{}\ its sizeable gaugino component that enhances
the couplings to $ \tilde{e}_{ \scriptscriptstyle  L,R} $ in
the $t$-channel. Scenario C is similar to B, but with a
larger $M_2$ ($M_2 = 1.5  M_{ \scriptscriptstyle  Z} $) that
gives a heavier $ \tilde{\chi}^{ \scriptscriptstyle  0}_1 $
and consequently a smaller cross section. The Scenario D
($\mu =3 M_{ \scriptscriptstyle  Z} $, $M_2 = 1.5 M_{
\scriptscriptstyle  Z} $) is almost symmetrical of A under
the transformation $\mu  \rightarrow  - \mu$ and gives both
$ \tilde{\chi}^{ \scriptscriptstyle  0}_1 $ and $
\tilde{\chi}^{ \scriptscriptstyle  0}_2 $ which are mostly
gauginos (with no really predominant $ \tilde{\gamma} $ or $
\tilde{Z} $ component) and lower production rates. Scenarios
H$^{\pm}$ are both in the HCS regions. Here both $
\tilde{\chi}^{ \scriptscriptstyle  0}_1 $ and $
\tilde{\chi}^{ \scriptscriptstyle  0}_2 $ are predominantly
higgsinos (of different kinds) and we get quite large cross
sections ($\sigma \approx 1.5$ pb). On the
contrary, the rates for $ \tilde{\chi}^{ \scriptscriptstyle
0}_1  \tilde{\chi}^{ \scriptscriptstyle  0}_1 $ production
are quite small, since for $ \tan \beta $ not far from 1,
one needs different higgsino components in the two produced
neutralinos in order to get a large coupling to
$ Z^{\scriptscriptstyle  0} $.

A similar analysis has been carried out in
\hbox{Table~}{}\ref{tab:scentgb4}, for $ \tan \beta =4$. The
corresponding six different scenarios E, F, G, J in NR and
H$^{\pm}$ in HCS are shown in
\hbox{Fig.~}{}\ref{fig:scentgb4}

We have also studied the $m_0$ and $ \sqrt{s} $ dependence
of $ e^{ \scriptscriptstyle  +}e^{ \scriptscriptstyle  -}
\rightarrow   \tilde{\chi}^{ \scriptscriptstyle  0}_1
\tilde{\chi}^{ \scriptscriptstyle  0}_2$ cross sections for
the six scenarios with $ \tan \beta =1.5$. In
\hbox{Fig.~}{}\ref{prodvsm0tgb1p5}, for $ \sqrt{s}  = 190$
GeV, we show the variation of cross sections with
$m_0$. One can see that Scenarios A,B,C and D, where the
$t$-channel amplitude is important, are the most affected by
the $m_0$ value. The maximal sensitivity is found in case A,
where the gaugino components are dominant in both $
\tilde{\chi}^{ \scriptscriptstyle  0}_1 $ and $
\tilde{\chi}^{ \scriptscriptstyle  0}_2 $. On the other
hand, production rates for scenarios H$^{\pm}$ are quite
insensitive to $m_0$, due to the $s$-channel dominance.

In  \hbox{Fig.~}{}\ref{prodvsrstgb1p5}, for
$m_0 =  M_{\scriptscriptstyle  Z} $, the $ \sqrt{s} $
dependence is studied around LEP2 energies.
Here too, one can notice the
different behaviour in various scenarios, due to the
relative importance of $t$- and $s$-channel contributions.
For each curve, the magnified symbols denote situations in
which the corresponding scenario is inside the Neutralino
Regions (\hbox{i.e.}{}, neutralino production is allowed,
but chargino production is not).

\section{Next-to-lightest neutralino decays}
\label{sec:decay}

In order to study possible signatures for $ \tilde{\chi}^{
\scriptscriptstyle  0}_1  \tilde{\chi}^{ \scriptscriptstyle
0}_2 $ production at LEP2, one has to analyse different
decay channels for the next-to-lightest neutralino. Indeed,
while $ \tilde{\chi}^{ \scriptscriptstyle  0}_1 $ will
always produce a considerable missing energy and missing
momentum signal, $ \tilde{\chi}^{ \scriptscriptstyle  0}_2 $
can give rise to a rich spectrum of final states
\cite{neutdec,raddec}. In
\hbox{Ref.~}{}\cite{ambmele}, a thorough study of all
possible $ \tilde{\chi}^{ \scriptscriptstyle  0}_2 $-decay
channels that are relevant at LEP2 has been performed. The
results of this analysis will be used in Section \ref{sec:totrates}
for the evaluation of total rates for different final states
corresponding to $ \tilde{\chi}^{ \scriptscriptstyle  0}_1
\tilde{\chi}^{ \scriptscriptstyle  0}_2 $ production at
LEP2. On the other hand, in this section, we report explicit
results for $ \tilde{\chi}^{ \scriptscriptstyle  0}_2 $
Branching Ratios (BR's), restricting ourselves to the
particular scenarios introduced in Section \ref{sec:prod}.

In the first column of  \hbox{Table~}{}\ref{tab:brtgb1p5},
all $ \tilde{\chi}^{ \scriptscriptstyle  0}_2 $ decays
allowed in the MSSM are listed. The first two channels refer
to the possibility for the $ \tilde{\chi}^{
\scriptscriptstyle  0}_2 $ to decay into either the lightest
scalar Higgs $ h^{ \scriptscriptstyle  0} $ or the
pseudoscalar Higgs $ A^{ \scriptscriptstyle  0} $
\cite{gunhab88}. For this reason, in
\hbox{Table~}{}\ref{tab:brtgb1p5}, we choose different
values for $ m_{A^{ \scriptscriptstyle  0}} $, that fix,
with $ \tan \beta $, the spectrum and couplings of the Higgs
sector. The following three channels include the main
three-body processes, that occur through the exchange of
either a $ Z^{ \scriptscriptstyle  0} $ gauge boson or a
scalar particle \cite{bartl88-91}. These latter decays may
occur in two steps, through production of a real scalar and
its subsequent decay into the corresponding fermion and a $
\tilde{\chi}^{ \scriptscriptstyle  0}_1 $. In fact, although
we are assuming $m_0 \ge  M_{ \scriptscriptstyle  Z} $ so
that $m_{\tilde{f}} >  \sqrt{s} /2$, when $m_0$ is close to
$ M_{ \scriptscriptstyle  Z} $, one or more sleptons
(usually the right-selectron and, sometimes, the sneutrino
that are the lightest sfermions,  \hbox{cfr.}{}\ Appendix
\ref{app:a}) may result lighter than the
$ \tilde{\chi}^{\scriptscriptstyle  0}_2 $. We do not consider
these situations separately and simply add these ``on-shell''
two-body contributions to the ``off-shell'' three-body ones.
Of course, the on-shell two-body decay considerably enhances
the corresponding BR. In the following two lines, we also
show BR's of $ \tilde{\chi}^{ \scriptscriptstyle  0}_2 $
decays into a light chargino plus either leptons or hadrons,
when allowed by phase-space. The last channel is the
one-loop radiative $ \tilde{\chi}^{ \scriptscriptstyle  0}_2
$ decay into a photon plus a $ \tilde{\chi}^{
\scriptscriptstyle  0}_1 $ \cite{raddec}, that gives rise in
the $ \tilde{\chi}^{ \scriptscriptstyle  0}_1
\tilde{\chi}^{ \scriptscriptstyle  0}_2 $ process to the
nice signature of one single photon production.
Unfortunately, this channel at LEP2 turns out to be in
general less important than at LEP1. Further details on
BR($\tilde{\chi}^{ \scriptscriptstyle  0}_2   \rightarrow
\tilde{\chi}^{ \scriptscriptstyle  0}_1 \gamma$) can be
found in  \hbox{Ref.~}{}\cite{ambmele}.

Let's start by considering situations where Higgses do not
contribute to two-body $ \tilde{\chi}^{ \scriptscriptstyle
0}_2 $ decays and the dominant channels are
$ \tilde{\chi}^{\scriptscriptstyle  0}_2   \rightarrow
\ell^{\scriptscriptstyle  +}\ell^{ \scriptscriptstyle  -}
\tilde{\chi}^{ \scriptscriptstyle  0}_1 $,
$\nu_{\scriptscriptstyle  \ell}
\bar{\nu}_{ \scriptscriptstyle \ell}
\tilde{\chi}^{ \scriptscriptstyle  0}_1 $,
$ q\bar{q}  \tilde{\chi}^{\scriptscriptstyle  0}_1 $
in the Neutralino Region scenarios.
In our framework, given $ \tan \beta $ (that is
equal to 1.5 in  \hbox{Table~}{}\ref{tab:brtgb1p5}) and
$M_2$, all sfermion masses are fixed by the value of $m_0$.
As a consequence, at fixed $m_0$, squark masses are in
general quite heavier than slepton masses and the $
\tilde{\chi}^{ \scriptscriptstyle  0}_2 $ decay into hadrons
coming from sfermion exchange are depressed with respect to
$ Z^{ \scriptscriptstyle  0} $-exchange contributions. In
scenarios A and D, where $|\mu|$ is relatively high and
gaugino components of $ \tilde{\chi}^{ \scriptscriptstyle
0}_1 $ and $ \tilde{\chi}^{ \scriptscriptstyle  0}_2 $ are
dominant (\hbox{cfr.}{}\
\hbox{Table~}{}\ref{tab:scentgb1p5}), only sfermion exchange
plays a r\^ole and leptonic channels almost saturate $
\tilde{\chi}^{ \scriptscriptstyle  0}_2 $ decays. Notice
that the numbers relative to the $ e^{ \scriptscriptstyle
+}e^{ \scriptscriptstyle  -}  \tilde{\chi}^{
\scriptscriptstyle  0}_1 $ channel refer to only one species
of charged leptons, while neutrino and quark channels are
summed up over all active flavours. Since the
right-selectron is lighter than $ \tilde{\chi}^{
\scriptscriptstyle  0}_2 $ in scenario A and not too much
heavier than $ \tilde{\chi}^{ \scriptscriptstyle  0}_2 $ in
scenario D (\hbox{cfr.}{}\
\hbox{Table~}{}\ref{tab:scentgb1p5}), the BR for $
\tilde{\chi}^{ \scriptscriptstyle  0}_2   \rightarrow
\tilde{\chi}^{ \scriptscriptstyle  0}_1  \ell^{
\scriptscriptstyle  +}\ell^{ \scriptscriptstyle  -} $ decay
turns out to be very large and of the order of 75\%, when
summed up over three charged lepton species. On the other
hand, scenarios B and C present mixed features and leptonic
channels are altogether comparable to the hadronic one.
Decays into chargino are not relevant in the above
scenarios, while the radiative $ \tilde{\chi}^{
\scriptscriptstyle  0}_2   \rightarrow   \tilde{\chi}^{
\scriptscriptstyle  0}_1 \gamma$ decay reaches a few per
cent of BR only in the B and C cases, where the total $
\tilde{\chi}^{ \scriptscriptstyle  0}_2 $ width is small due
to the mixed gaugino-higgsino nature of $ \tilde{\chi}^{
\scriptscriptstyle  0}_1 $ and $ \tilde{\chi}^{
\scriptscriptstyle  0}_2 $. As for scenario H$^-$, where
higgsino components are dominant, the $ Z^{
\scriptscriptstyle  0} $ exchange saturates $ \tilde{\chi}^{
\scriptscriptstyle  0}_2 $ decays. Hence, BR's for various
channels closely reflect the branching ratios for $ Z^{
\scriptscriptstyle  0}   \rightarrow   f\bar{f}$. In
scenario H$^+$, the $ Z^{ \scriptscriptstyle  0} $-channel
dominance is less pronounced. Also, there is a considerable
BR for channels with a $ \tilde{\chi}^{ \scriptscriptstyle
\pm}_1 $ in the final state, that gives rise to $
\tilde{\chi}^{ \scriptscriptstyle  0}_2 $ cascade decays.

When Higgs masses are sufficiently light to allow the decays
$ \tilde{\chi}^{ \scriptscriptstyle  0}_2   \rightarrow
\tilde{\chi}^{ \scriptscriptstyle  0}_1  h^{
\scriptscriptstyle  0} $, $ \tilde{\chi}^{
\scriptscriptstyle  0}_1  A^{ \scriptscriptstyle  0} $,
these channels are always important. For instance, in the
case A with $ m_{A^{ \scriptscriptstyle  0}}  =  M_{
\scriptscriptstyle  Z} $, there is a $98\%$ probability for
$ \tilde{\chi}^{ \scriptscriptstyle  0}_2   \rightarrow
\tilde{\chi}^{ \scriptscriptstyle  0}_1  h^{
\scriptscriptstyle  0} $ (see
\hbox{Table~}{}\ref{tab:brtgb1p5}). In the case H$^+$ with
$ m_{A^{ \scriptscriptstyle  0}}  =  M_{ \scriptscriptstyle
Z} /2$, the BR for $ \tilde{\chi}^{ \scriptscriptstyle  0}_2
  \rightarrow   \tilde{\chi}^{ \scriptscriptstyle  0}_1  A^{
\scriptscriptstyle  0} $ is about $21\%$. In general, Higgs
channels will give rise to an enhanced hadronic signal
coming from $ h^{ \scriptscriptstyle  0} ,  A^{
\scriptscriptstyle  0}    \rightarrow    b\bar{b} $. For $
\tilde{\chi}^{ \scriptscriptstyle  0}_2 $ decays into
Higgses, the $ h^{ \scriptscriptstyle  0} $ mass and
couplings have been computed taking into account the leading
one-loop corrections due to top-stop contributions in the
approximation of degenerate stop masses (\hbox{cfr.}{}\
Appendix \ref{app:a}).

A rather different picture emerges for $ \tan \beta  = 4$
(\hbox{Table~}{}\ref{tab:brtgb4}). First of all, when moving
up from $ \tan \beta  = 1$, it is harder and harder to find
scenarios where one of the two lightest neutralinos is
almost a pure gaugino and the other one is almost a pure
higgsino in the parameter space relevant for LEP2.
Consequently, the tree-level decays of $ \tilde{\chi}^{
\scriptscriptstyle  0}_2 $ are never much depressed and the
BR for the radiative channel $ \tilde{\chi}^{
\scriptscriptstyle  0}_2   \rightarrow   \tilde{\chi}^{
\scriptscriptstyle  0}_1 \gamma$ can be at most a few per
mil in the Neutralino Region. Furthermore, varying $ \tan
\beta $ changes both the gaugino-higgsino composition of
$\tilde{\chi}^{\scriptscriptstyle 0}_{1,2} $
and the scalar mass spectrum. One of the main
effects of that is the relative decreasing of the sneutrino
mass with respect to selectron masses
(\hbox{cfr.}{}\ Appendix \ref{app:a}).

In  \hbox{Table~}{}\ref{tab:brtgb4}, one can see that in
scenarios E and J with large $|\mu|$ and for heavy Higgses,
the BR for $ \tilde{\chi}^{ \scriptscriptstyle  0}_2
\rightarrow   \tilde{\chi}^{ \scriptscriptstyle  0}_1  \nu_{
\scriptscriptstyle  \ell}  \bar{\nu}_{ \scriptscriptstyle
\ell} $ turns out to be considerably enhanced and decreases
the visible fraction of $ \tilde{\chi}^{ \scriptscriptstyle
0}_2 $ decays. As for decays into real Higgses, one should
take into account that the $ h^{ \scriptscriptstyle  0} $
mass increases with $ \tan \beta $ at fixed $ m_{A^{
\scriptscriptstyle  0}} $ (\hbox{cfr.}{}\
\hbox{Eqs.~}{}(\ref{eq:hmass})). For this reason, in
\hbox{Table~}{}\ref{tab:brtgb4}, lower values of $ m_{A^{
\scriptscriptstyle  0}} $ with respect to
\hbox{Table~}{}\ref{tab:brtgb1p5} have been chosen to
characterize scenarios with allowed and not-allowed $
\tilde{\chi}^{ \scriptscriptstyle  0}_2    \rightarrow
\tilde{\chi}^{ \scriptscriptstyle  0}_1  h^{
\scriptscriptstyle  0} $, $ \tilde{\chi}^{
\scriptscriptstyle  0}_1  A^{ \scriptscriptstyle  0} $
decays. Here again, when permitted by phase space, the
two-body decay into Higgses almost saturates the BR.

As in the case $ \tan \beta  = 1.5$, the detailed features
of each decay BR in  \hbox{Table~}{}\ref{tab:brtgb4} can be
understood by considering the physical composition of
neutralinos given in  \hbox{Table~}{}\ref{tab:scentgb4}, for
scenarios with $ \tan \beta  = 4$.

\section{Total rates for different
$\protect \tilde{\chi}^{ \scriptscriptstyle  0}_1
\protect \tilde{\chi}^{ \scriptscriptstyle  0}_2 $
signatures}
\label{sec:totrates}

In this section, we will show results on total rates
corresponding to different signatures coming from the
process $ e^{ \scriptscriptstyle  +}e^{ \scriptscriptstyle
-}    \rightarrow   \tilde{\chi}^{ \scriptscriptstyle 0}_1
\tilde{\chi}^{ \scriptscriptstyle  0}_2$, with particular
emphasis on the Neutralino Regions. Various rates are
obtained by multiplying cross sections with BR's for
different decay channels of $ \tilde{\chi}^{
\scriptscriptstyle  0}_2 $ at fixed values of $M_2$, $\mu$,
$m_0$, $ \tan \beta $ and $ m_{A^{ \scriptscriptstyle  0}}
$.

While $ \tilde{\chi}^{ \scriptscriptstyle  0}_1 $ always
produces a large missing energy and missing momentum signal
in the final state, each $ \tilde{\chi}^{ \scriptscriptstyle
 0}_2 $ decay channel contributes to a different signature.
The most interesting signatures correspond to some visible
either leptonic or hadronic (less often mixed) signal
concentrated in the opposite side with respect to the $
\tilde{\chi}^{ \scriptscriptstyle  0}_1 $ direction. In our
analysis, we neglect hadronization effects and in general we
assume that each quark gives rise to a jet in the final
state. \\ We now proceed to listing all the possible
signatures corresponding to $ e^{ \scriptscriptstyle  +}e^{
\scriptscriptstyle  -}    \rightarrow \tilde{\chi}^{
\scriptscriptstyle  0}_1 \tilde{\chi}^{ \scriptscriptstyle
0}_2$:
\begin{description}
\item[i) $ \ell^{
\scriptscriptstyle  +}\ell^{ \scriptscriptstyle  -}  +
\not\!\! E $], coming in general from $ \tilde{\chi}^{
\scriptscriptstyle  0}_2   \rightarrow   \tilde{\chi}^{
\scriptscriptstyle  0}_1  \ell^{ \scriptscriptstyle
+}\ell^{ \scriptscriptstyle  -} $ ($\ell = e, \mu$). The
same signature, but with softer leptons, is obtained for
particular scenarios where cascade decays of $
\tilde{\chi}^{ \scriptscriptstyle  0}_2 $ through a lighter
$ \tilde{\chi}^{ \scriptscriptstyle  \pm}_1 $ are allowed:
\begin{equation}
\begin{array}{r c l}
e^{ \scriptscriptstyle  +}e^{ \scriptscriptstyle  -} &
\rightarrow &
\tilde{\chi}^{ \scriptscriptstyle 0}_1 \;
\tilde{\chi}^{ \scriptscriptstyle  0}_2 \\
& & \phantom{\; \tilde{\chi}^{ \scriptscriptstyle 0}_1}
\:\raisebox{1.3ex}{\rlap{$\vert$}}\!\rightarrow
\ell^{\scriptscriptstyle  \pm}  \nu_{ \scriptscriptstyle  \ell}
\tilde{\chi}^{ \scriptscriptstyle \mp} \\
& &
\phantom{\tilde{\chi}^{ \scriptscriptstyle 0}_1}
\phantom{\rightarrow}
\phantom{\ell^{\scriptscriptstyle  \pm}
\nu_{ \scriptscriptstyle  \ell}}
\phantom{\tilde{\chi}^{ \scriptscriptstyle \mp}}
\! \:\raisebox{1.3ex}{\rlap{$\vert$}}\!\rightarrow
\ell^{ \scriptscriptstyle \mp}
\nu_{ \scriptscriptstyle  \ell}
\tilde{\chi}^{
\scriptscriptstyle  0}_1 \; .
\end{array}
\label{nntonllvv}
\end{equation}
In general, we will
see that these cascade decays are relevant in regions of the
$(\mu, M_2)$ plane where also chargino-pair production can
occur. By the way, the process (\ref{nntonllvv}) will give
rise, with twice the BR for $ \tilde{\chi}^{
\scriptscriptstyle  0}_2   \rightarrow   \ell^{
\scriptscriptstyle  +}\ell^{ \scriptscriptstyle  -}  +
\tilde{\chi}^{ \scriptscriptstyle  0}_1 $, to the signature
$ \ell^{ \scriptscriptstyle  +} \ell^{ \scriptscriptstyle
\prime -} +  \not\!\! E $ with a pair of leptons of
different flavours. We also notice once more that our choice
of $m_0$ (which prevents pair production of sleptons at
LEP2), hinders in general the two-body decay $
\tilde{\chi}^{ \scriptscriptstyle  0}_2   \rightarrow
\tilde{\ell}^{ \scriptscriptstyle  \pm}_{ \scriptscriptstyle
 L,R}  \ell^{ \scriptscriptstyle  \mp} $. Nevertheless,
there are particular choices of {\sc SuSy} parameters (
\hbox{e.g.}{}\ scenario A) that allow the $ \tilde{\chi}^{
\scriptscriptstyle  0}_2 $ decay into a real charged slepton
that is too heavy to be pair-produced at LEP2. The same can
happen with the $ \tilde{\chi}^{ \scriptscriptstyle  0}_2
\rightarrow   \tilde{\nu}_{ \scriptscriptstyle  \ell,L}
\bar{\nu}_{ \scriptscriptstyle  \ell} $, $
\bar{\tilde{\nu}}_{ \scriptscriptstyle  \ell,L}  \nu_{
\scriptscriptstyle  \ell} $ channel, but not for $
\tilde{\chi}^{ \scriptscriptstyle  0}_2   \rightarrow
\tilde{q}_{ \scriptscriptstyle  L,R}  \bar{q}$, $
\bar{\tilde{q}}_{ \scriptscriptstyle  L,R}  q$, since
squarks are generally quite heavier than sleptons
(\hbox{cfr.}{}\ Appendix \ref{app:a}). This is also supported
by the stronger limits found by CDF on $m_{\tilde{q}_{L,R}}$,
with respect to the LEP limits on $m_{\tilde{\ell}_{L,R}}$
\cite{pdg}. Correspondingly, one can have situations in
which the leptonic decays almost saturate the $
\tilde{\chi}^{ \scriptscriptstyle  0}_2 $ width. In
particular, if only $ \tilde{\chi}^{ \scriptscriptstyle
0}_2   \rightarrow   \tilde{\ell}^{ \scriptscriptstyle
\pm}_{ \scriptscriptstyle  R}   \ell^{ \scriptscriptstyle
\mp} $ is allowed, as in scenario A ($ \tilde{\ell}_{
\scriptscriptstyle  R} $ is in general lighter than $
\tilde{\ell}_{ \scriptscriptstyle  L} $ and $ \tilde{\nu}_{
\scriptscriptstyle  \ell,L} $), the $ \ell^{
\scriptscriptstyle  +}\ell^{ \scriptscriptstyle  -}  +
\not\!\! E $ signal is at least $25\%$ of the total for each
lepton flavour.
\item[ii) 2j $+  \not\!\! E $], arising from
the decay $ \tilde{\chi}^{ \scriptscriptstyle  0}_2
\rightarrow   q\bar{q}  \tilde{\chi}^{ \scriptscriptstyle
0}_1 $. Our results are always summed up over five quark
flavours (in the massless-quark approximation). When allowed
by phase space, also the two-body decays $ \tilde{\chi}^{
\scriptscriptstyle  0}_2   \rightarrow   \tilde{\chi}^{
\scriptscriptstyle  0}_1  h^{ \scriptscriptstyle  0} $, $
\tilde{\chi}^{ \scriptscriptstyle  0}_1  A^{
\scriptscriptstyle  0} $ enter this class, due to the
subsequent $ h^{ \scriptscriptstyle  0} , A^{
\scriptscriptstyle  0}   \rightarrow   b\bar{b} $. In our
analysis, we will sum this contribution, when present, to
the direct 2-jet signal.
\item[iii) $\gamma +  \not\!\! E
$], coming from the one-loop decay $ \tilde{\chi}^{
\scriptscriptstyle  0}_2   \rightarrow   \tilde{\chi}^{
\scriptscriptstyle  0}_1 \gamma$.
\item[iv) 4j $+  \not\!\!
E $]. This arises from the cascade decay $ \tilde{\chi}^{
\scriptscriptstyle  0}_2   \rightarrow   \tilde{\chi}^{
\scriptscriptstyle  \pm}_1 (  \rightarrow   q_1\bar{q}_1^{
\scriptscriptstyle \prime} \tilde{\chi}^{ \scriptscriptstyle
 0}_1 )q_2\bar{q}_2^{ \scriptscriptstyle  \prime}$,
similarly to (\ref{nntonllvv}).
\item[v) $ \ell^{\scriptscriptstyle  \pm}  +$ 2j $+
\not\!\! E $], still coming from
$ \tilde{\chi}^{ \scriptscriptstyle  0}_2 $
cascade decays $ \tilde{\chi}^{ \scriptscriptstyle  0}_2
\rightarrow   \tilde{\chi}^{ \scriptscriptstyle  \mp}_1
(\rightarrow q\bar{q}^{ \scriptscriptstyle  \prime}
\tilde{\chi}^{ \scriptscriptstyle  0}_1 )
\ell^{\scriptscriptstyle  \pm}
\nu_{ \scriptscriptstyle  \ell} $ \\
or $ \tilde{\chi}^{ \scriptscriptstyle  0}_2   \rightarrow
\tilde{\chi}^{ \scriptscriptstyle  \pm}_1 (  \rightarrow
\ell^{ \scriptscriptstyle  \pm}  \nu_{ \scriptscriptstyle
\ell}  \tilde{\chi}^{ \scriptscriptstyle  0}_1 )
q\bar{q}^{\scriptscriptstyle \prime}$.
\item[vi) invisible]. This
arises from channels that have only neutrinos and $
\tilde{\chi}^{ \scriptscriptstyle  0}_1 $'s in the final
state, that is $ \tilde{\chi}^{ \scriptscriptstyle  0}_2
\rightarrow   \nu_{ \scriptscriptstyle  \ell}
\bar{\nu}_{\scriptscriptstyle  \ell}
\tilde{\chi}^{ \scriptscriptstyle 0}_1 $. Of course, this
is the least interesting signature,
that, at LEP2, can be seen only if there is some detected
radiation emitted from the initial state. This case is
unfavoured due to both a lower cross section corresponding
to the initial state photon radiation and a smaller
effective  \hbox{c.m.}{}\ energy left for the $
\tilde{\chi}^{ \scriptscriptstyle  0}_1  \tilde{\chi}^{
\scriptscriptstyle  0}_2 $ production. The latter effect
takes the available parameter space back to the region
covered by chargino search.
\item[vii) $ \tau^{
\scriptscriptstyle  +}\tau^{ \scriptscriptstyle  -}  +
\not\!\! E $]. Most of the times, this signal arises from
the direct decay $ \tilde{\chi}^{ \scriptscriptstyle  0}_2
\rightarrow   \tau^{ \scriptscriptstyle  +}\tau^{
\scriptscriptstyle  -}  \tilde{\chi}^{ \scriptscriptstyle
0}_1 $, analogously to the events of class {\bf i} above. A
non-negligible contribution to this channel comes also from
direct decays $ \tilde{\chi}^{ \scriptscriptstyle  0}_2
\rightarrow   \tilde{\chi}^{ \scriptscriptstyle  0}_1  h^{
\scriptscriptstyle  0} $, $ \tilde{\chi}^{
\scriptscriptstyle  0}_1  A^{ \scriptscriptstyle  0} $ when
allowed by phase space. Tau production gives rise to various
signatures, where, in general, the visible energy is lower
than for previous cases, due to the presence of at least two
neutrinos in the final state. One can have: a $ \ell^{
\scriptscriptstyle  +} \ell^{ \scriptscriptstyle  (\prime)
-} +  \not\!\! E $ signature, with a BR of about $12\%$,
hadrons $+  \not\!\! E $  with a $41\%$ probability and, in
the remaining cases, $ \ell^{ \scriptscriptstyle  \pm}  +$
hadrons $+  \not\!\! E $. In the following analysis, we will
keep separate the contributions to hadronic and ($e$-$\mu$)
leptonic signals coming from $\tau$ decays from the main
ones described in classes {\bf i-v}.
\end{description}

One should keep in mind that, in order to get a detectable
signal arising from the above decay channels, the mass
difference between $ \tilde{\chi}^{ \scriptscriptstyle  0}_2
$ and $ \tilde{\chi}^{ \scriptscriptstyle  0}_1 $ must be
sizeable. Indeed, jets, leptons and photons should have
enough visible energy. We have checked that this condition
is in general fulfilled in the region of the {\sc SuSy} parameter
space not excluded by LEP1. A few exceptions occur in very
limited regions for $ \tan \beta \simeq 1$. This aspect can
be more dramatic for jets and/or leptons arising from $
\tilde{\chi}^{ \scriptscriptstyle  0}_2 $ cascade decays
through a light chargino. In this case, at least one of the
two differences $( m_{\tilde{\chi}^0_2} -
m_{\tilde{\chi}^{\pm}_1} )$ and $( m_{\tilde{\chi}^{\pm}_1}
- m_{\tilde{\chi}^0_1} )$ should be sizeable. We will
explicitly show unfavourable regions (that, in most cases,
do not overlap with the parameter space explorable at LEP2)
in the following discussion.

In  \hbox{Figs.~}{}\ref{eveetgb15ma273},
\ref{ev2jtgb15ma273} and \ref{evphtgb15ma273}, we show the
contour plots in the $(\mu, M_2)$ plane for total rates (in
fb) corresponding to signatures {\bf i-iii}, coming from
direct (and, if relevant, from cascade)
$ \tilde{\chi}^{\scriptscriptstyle  0}_2 $ decays when
$ \tilde{\chi}^{\scriptscriptstyle  0}_2 $ two-body decays
into Higgses are not allowed.
We set $ \sqrt{s}  = 190$ GeV, $m_0=M_{\scriptscriptstyle Z}$
and  $ \tan \beta  = 1.5$. The precise value of
$ m_{A^{ \scriptscriptstyle  0}}  = 3 M_{\scriptscriptstyle Z}$,
besides pushing Higgs masses above threshold for
$ \tilde{\chi}^{ \scriptscriptstyle  0}_2  \rightarrow
\tilde{\chi}^{\scriptscriptstyle 0}_1  h^{\scriptscriptstyle 0}$,
$\tilde{\chi}^{\scriptscriptstyle 0}_1  A^{\scriptscriptstyle 0}$,
is relevant for the single photon signal, since charged Higgses
enter into loops for the radiative
$\tilde{\chi}^{\scriptscriptstyle  0}_2   \rightarrow
\tilde{\chi}^{\scriptscriptstyle  0}_1 \gamma$ decay (in general larger
$m_{H^{ \scriptscriptstyle  \pm}}$ gives higher rates for
this signal,  \hbox{cfr.}{}\ \hbox{Table~}{}\ref{tab:brtgb1p5}).
With an integrated luminosity of 500 pb$^{-1}$, it is straightforward
to get the expected number of events at LEP2 by halving the numbers
shown in the figures.

The thick bold line in  \hbox{Figs.~}{}\ref{eveetgb15ma273},
\ref{ev2jtgb15ma273} and \ref{evphtgb15ma273} binds the
small area where $( m_{\tilde{\chi}^0_2} -
m_{\tilde{\chi}^0_1} ) < 10$ GeV. This may help in
selecting regions where the final particles are actually
visible in all direct decays.

As for the $ e^{ \scriptscriptstyle  +}e^{
\scriptscriptstyle  -}  +  \not\!\! E $ signal, we can note
in  \hbox{Fig.~}{}\ref{eveetgb15ma273} that in the NR$^+$
one gets up to 60 fb (equivalent to 30 events), mainly due
to the presence of a light selectron, while in the NR$^+$,
rates reach at most 20 fb. Comparable rates are obtained in
the High Cross Section regions HCS$^{\pm}$, due to the
dominance of $ Z^{ \scriptscriptstyle  0} $ channels, that
keep relevant leptonic $ \tilde{\chi}^{ \scriptscriptstyle
0}_2 $-decay BR's down to values of the order of
BR($ Z^{\scriptscriptstyle  0}   \rightarrow
e^{\scriptscriptstyle  +}e^{ \scriptscriptstyle  -} $).

A quite less favourable situation is found for the 2j $+
\not\!\! E $ signature for the same set of parameters. In
\hbox{Fig.~}{}\ref{ev2jtgb15ma273}, we can see that in most
of the Neutralino Regions the rate for this signal is too
low to be detected at LEP2. In particular for $|\mu|>2  M_{
\scriptscriptstyle  Z} $ in the NR's, one finds less than 5
fb, while some signal can be detected in the NR$^+$ for $-2
 M_{ \scriptscriptstyle  Z}  < \mu
\;\raisebox{-.5ex}{\rlap{$\sim$}} \raisebox{.5ex}{$<$}\;
M_{ \scriptscriptstyle  Z} $. In this region, where $ Z^{
\scriptscriptstyle  0} $-channel $ \tilde{\chi}^{
\scriptscriptstyle  0}_2 $ decays (which are not depressed
by squark masses) are important due to the quite large
higgsino component of $ \tilde{\chi}^{ \scriptscriptstyle
0}_2 $ (\hbox{cfr.}{}\
\hbox{Table~}{}\ref{tab:scentgb1p5}) rates up to 100 fb can
be reached. Differently from the leptonic signature of
\hbox{Fig.~}{}\ref{eveetgb15ma273}, in the HCS regions the
large value of BR($ Z^{ \scriptscriptstyle  0}   \rightarrow
q\bar{q} $) gives rise to total rates of the order of 1 pb.

In  \hbox{Fig.~}{}\ref{evphtgb15ma273}, the single photon
rate coming from $ \tilde{\chi}^{ \scriptscriptstyle  0}_2
\rightarrow   \tilde{\chi}^{ \scriptscriptstyle  0}_1
\gamma$ is shown. No signal is obtained for $\mu>0$, while
in the $\mu<0$ half-plane one can reach at most about 100 fb
in the region covered by chargino search. Restricting
ourselves to the NR$^+$, rates up to about 50 fb are found
in the area close to $\mu = - M_{ \scriptscriptstyle  Z} $,
just on the edge of the chargino-pair production region.
Unfortunately, in the regions where the rate for the $\gamma
+  \not\!\! E $ signal exceeds 50 fb, the emitted photon is
likely to be quite soft, due to the small difference between
neutralino masses ($ \;\raisebox{-.5ex}{\rlap{$\sim$}}
\raisebox{.5ex}{$<$}\;  10$ GeV, see the thick bold
line in  \hbox{Fig.~}{}\ref{evphtgb15ma273}). In the HCS
regions, some signal is found only for $\mu<0$ and very
large $M_2$ values.

The total  rate coming from  all visible $ \tilde{\chi}^{
\scriptscriptstyle  0}_2 $-decay channels (including cascade
decays, see below) is reported in
\hbox{Fig.~}{}\ref{evvistgb15ma273}, where the same set of
{\sc SuSy} parameters as for
\hbox{Fig.~}{}\ref{eveetgb15ma273}--\ref{evphtgb15ma273} has
been chosen. By comparing this figure with
\hbox{Fig.~}{}\ref{n1n2prodm91tgb15rs190} for total cross
section, one can assess rates from invisible $
\tilde{\chi}^{ \scriptscriptstyle  0}_2 $ decays, that for $
\tan \beta  = 1.5$ are typically 10-20\% of the total.

The effect on the hadronic signal of assuming a lighter
Higgs spectrum and in particular of allowing the decay $
\tilde{\chi}^{ \scriptscriptstyle  0}_2   \rightarrow   h^{
\scriptscriptstyle  0}  \tilde{\chi}^{ \scriptscriptstyle
0}_1 $ is shown in  \hbox{Fig.~}{}\ref{ev2jtgb15ma91}, that
gives the 2j $+  \not\!\! E $ rates for $ m_{A^{
\scriptscriptstyle  0}}  =  M_{ \scriptscriptstyle  Z} $
(the values of remaining parameters are the same as for
\hbox{Fig.~}{}\ref{ev2jtgb15ma273}). By comparing
\hbox{Fig.~}{}\ref{ev2jtgb15ma91} and \ref{ev2jtgb15ma273},
one finds a remarkable enhancement of the hadronic signal in
the $(\mu,M_2)$ parameter space where the $ \tilde{\chi}^{
\scriptscriptstyle  0}_2 $-$ \tilde{\chi}^{
\scriptscriptstyle  0}_1 $ mass difference is larger than $
m_{h^{ \scriptscriptstyle  0}} $. This mainly happens for
$\mu < - M_{ \scriptscriptstyle  Z} $ in the Neutralino
Regions. Hence, the bulk of the leptonic signal in the
NR$^+$ is substituted by the 2-jet signal (where the two
jets are predominantly $b$-quark jets coming from $ h^{
\scriptscriptstyle  0}   \rightarrow   b\bar{b} $).

The rate for 4j $+  \not\!\! E $ arising from the cascade
decay $ \tilde{\chi}^{ \scriptscriptstyle  0}_2
\rightarrow   \tilde{\chi}^{ \scriptscriptstyle  \pm}_1
( \rightarrow q_1\bar{q}_1^{ \scriptscriptstyle \prime}
\tilde{\chi}^{ \scriptscriptstyle  0}_1 )
q_2\bar{q}_2^{\scriptscriptstyle  \prime}$ is shown in
\hbox{Fig.~}{}\ref{ev4jtgb15ma273}, for $m_0 =
M_{\scriptscriptstyle  Z} $ and $ \tan \beta  = 1.5$. The
heavy Higgs case is considered ($ m_{A^{ \scriptscriptstyle  0}}
= 3  M_{ \scriptscriptstyle  Z} $). One can see that a
considerable signal is found for positive $\mu$, but not in
regions not covered by direct chargino search. In the area
not excluded by LEP1, one gets rates up to about 200 fb.

In  \hbox{Fig.~}{}\ref{ev4jtgb15ma273}, outside the bold
dashed line, one has $( m_{\tilde{\chi}^0_2}  -
m_{\tilde{\chi}^{\pm}_1} ) < 10$ GeV, while the
region where $( m_{\tilde{\chi}^{\pm}_1}  -
m_{\tilde{\chi}^0_1} ) < 10$ GeV is completely
contained in the area excluded by LEP1. Therefore, one can
hope to have a visible 4-jet signal in most of the
large-rate region, while at least 2 jets should be always
detectable. An analogous conclusion applies to the other
cascade-decay signatures.

The mixed semileptonic signature $e^{\scriptscriptstyle +}
+$ 2j $+  \not\!\! E $, still coming from
$\tilde{\chi}^{ \scriptscriptstyle  0}_2 $ cascade decays
mediated by a light chargino, is studied in
\hbox{Fig.~}{}\ref{eve2jtgb15ma273}, for the same set of
$m_0$, $ m_{A^{ \scriptscriptstyle  0}} $ and $ \tan \beta $
values. Rates refer to a positron in the final state and
must be doubled when summing up over lepton charges. The
picture in the $(\mu,M_2)$ plane is similar to the previous
4j $+  \not\!\! E $ one, with lower rates mainly due to the
single leptonic flavour considered. A similar  behaviour is
found also for the $ e^{ \scriptscriptstyle  +}   \mu^{
\scriptscriptstyle  -}  +  \not\!\! E $ rates, which are
however further reduced and at the edge of detectability.

Up to now, we considered the case $ \tan \beta  = 1.5$.
Increasing $ \tan \beta $ value in general  makes the
situation worse due to the combined effects of the shift of
the NR area (that for larger $ \tan \beta $ value tends to
be more symmetric with respect to the inversion of the $\mu$
sign) and of the reduction of $ \tilde{\chi}^{
\scriptscriptstyle  0}_2 $ BR's for visible decays. In
\hbox{Fig.~}{}\ref{evvistgb4ma91}, the rates for the total
visible signal are shown for $m_0 =  M_{ \scriptscriptstyle
Z} $, $ m_{A^{ \scriptscriptstyle  0}}  =  M_{
\scriptscriptstyle  Z} $ and $ \tan \beta  = 4$. The
relatively light $ m_{A^{ \scriptscriptstyle  0}} $ in this
case does not give rise to direct $ \tilde{\chi}^{
\scriptscriptstyle  0}_2   \rightarrow   \tilde{\chi}^{
\scriptscriptstyle  0}_1  h^{ \scriptscriptstyle  0} $
decays, due to the strong dependence of $ m_{h^{
\scriptscriptstyle  0}} $ on $ \tan \beta $ (\hbox{cfr.}{}\
 \hbox{Table~}{}\ref{tab:brtgb4}). One can see that the
Neutralino Region rates are lower than for $ \tan \beta  =
1.5$ (\hbox{cfr.}{}\  \hbox{Fig.~}{}\ref{evvistgb15ma273}),
especially in the NR$^+$.

This trend is kept for even higher values of $ \tan \beta $.
For instance, in  \hbox{Fig.~}{}\ref{evvistgb30ma273}, the
visible signal for $ \tan \beta  = 30$ and $m_0 =  M_{
\scriptscriptstyle  Z} $, is shown. Here, the NR$^+$ is
further reduced and the visible rate is greater than about
100 fb only outside the Neutralino Regions. The same pattern
is observed for the
$ e^{ \scriptscriptstyle  +}e^{\scriptscriptstyle  -}  +
\not\!\! E $ and 2 jets $+ \not\!\! E $ rates.
As for the $\gamma +  \not\!\! E $
signal, its cross section never exceeds a few fb's outside
the region covered by LEP1, while rates for the 4 jets $+
\not\!\! E $ cascade decay (that are similar to the
$ e^{\scriptscriptstyle  +}  +$ 2 jets $+  \not\!\! E $ ones)
are shown in  \hbox{Fig.~}{}\ref{ev4jtgb30ma273}.

As for the detectability of jets, leptons and photons at
high $ \tan \beta $, we have checked that, for $ \tan \beta
 \;\raisebox{-.5ex}{\rlap{$\sim$}} \raisebox{.5ex}{$>$}\;
4$, neutralino- and chargino-mass differences are always
sufficient to provide enough energy to the final particles
in regions relevant at LEP2.

Finally, in  \hbox{Tables~}{}\ref{tab:evtgb1p5} and
\ref{tab:evtgb4} total rates corresponding to all possible
signatures coming from $ e^{ \scriptscriptstyle  +}e^{
\scriptscriptstyle  -}    \rightarrow \tilde{\chi}^{
\scriptscriptstyle  0}_1 \tilde{\chi}^{ \scriptscriptstyle
0}_2$ at LEP2 are shown for the scenarios defined above with
$ \tan \beta  = 1.5$ and $ \tan \beta  = 4$, respectively.
Note that in these tables the rates relative to each
signature include all possible contributions. For instance,
$ \tau^{ \scriptscriptstyle  +}  \tau^{ \scriptscriptstyle
-}  +  \not\!\! E $ includes both the direct $
\tilde{\chi}^{ \scriptscriptstyle  0}_2    \rightarrow
\tau^{ \scriptscriptstyle  +}\tau^{ \scriptscriptstyle  -}
\tilde{\chi}^{ \scriptscriptstyle  0}_1 $ decay and the two
processes $ \tilde{\chi}^{ \scriptscriptstyle  0}_2
\rightarrow  h^{ \scriptscriptstyle  0} , A^{
\scriptscriptstyle  0} (  \rightarrow   \tau^{
\scriptscriptstyle  +}\tau^{ \scriptscriptstyle  -} )
\tilde{\chi}^{ \scriptscriptstyle  0}_1 $ and $
\tilde{\chi}^{ \scriptscriptstyle  0}_2   \rightarrow
\tilde{\chi}^{ \scriptscriptstyle  \pm}_1 (  \rightarrow
\tau^{ \scriptscriptstyle  \pm}  \nu_{ \scriptscriptstyle
\tau} )  \tau^{ \scriptscriptstyle  \mp}  \nu_{
\scriptscriptstyle  \tau}  \tilde{\chi}^{ \scriptscriptstyle
 0}_1 $.

We now make some comments on specific Neutralino-Region
scenarios. In scenario A, for heavy Higgses, one has a
considerable leptonic signal, corresponding to about 40 $
e^{ \scriptscriptstyle  +}e^{ \scriptscriptstyle  -} ,
\mu^{ \scriptscriptstyle  +}\mu^{ \scriptscriptstyle  -} +
\not\!\! E $ events for 500 pb$^{-1}$. For light Higgses,
this is replaced by an even larger hadronic ($ b\bar{b}  +
\not\!\! E $) signal. In scenario B, independently from
Higgses, the bulk of the total visible signal (that is about
105 fb) comes from the hadronic signature, mostly due to
light quarks. Rather lower rates correspond to scenario C,
where again most of the signal corresponds to hadronic final
states. Even lower rates correspond to scenario D, where all
the visible signal (about 45 fb) comes from charged lepton
pairs. As for the single photon signal, we find at most a
few fb's in NR$^+$.

A less favourable situation is found for the Neutralino
Region scenarios with $ \tan \beta  = 4$
(\hbox{Table~}{}\ref{tab:evtgb4}). Here, unless Higgses are
light enough to allow direct two-body decays, total visible
rates never exceed 40 fb. For light Higgses, one can reach
in scenario E a total visible signal of about 111 fb.

As for the High Cross Section regions, while the bulk of
visible rates corresponds to 2 jets $+  \not\!\! E $ signal,
there are non-negligible rates even for more interesting
signatures coming from cascade $ \tilde{\chi}^{
\scriptscriptstyle  0}_2 $ decays into charginos. For
instance, in the H$^+$ scenario for $ \tan \beta  = 1.5$,
one finds about 162 fb for the signal 4 jets $+  \not\!\! E
$ and about 109 fb (4 times 27.2) for the signal $ e^{
\scriptscriptstyle  \pm}  \ ( \mu^{ \scriptscriptstyle  \pm}
) + 2j +  \not\!\! E $.

In the first phase of running at LEP2, the  \hbox{c.m.}{}\
energy will be slightly lower (\hbox{i.e.}{}, $ \sqrt{s}  =
175$ GeV) than the one assumed here. Small
differences are expected in this case. The general trend of
variation can be inferred by comparing
\hbox{Fig.~}{}\ref{evvistgb15ma273}, at $ \sqrt{s}  = 190$
GeV, with  \hbox{Fig.~}{}\ref{evvisrs175}, at $
\sqrt{s}  = 175$ GeV, for the visible cross
section. On the one hand, there is a small reduction of the
explorable region in the {\sc SuSy} parameter space, due to the
smaller available phase space at $ \sqrt{s}  = 175$ GeV (\hbox{cfr.}{}\
\hbox{Fig.~}{}\ref{prodvsrstgb1p5}). As a consequence, one
can observe that the relative importance of the Neutralino
Region with respect to the chargino region is slightly
increased. On the other hand, in HCS regions, where
$s$-channel $ Z^{ \scriptscriptstyle  0} $-exchange
dominates, cross sections generally grow by about 20\% at $
\sqrt{s}  = 175$ GeV.

\vspace{0.5cm}

In conclusion, we have found that neutralino production
through the channel $ e^{ \scriptscriptstyle  +}
e^{ \scriptscriptstyle  -}    \rightarrow
\tilde{\chi}^{ \scriptscriptstyle  0}_1
\tilde{\chi}^{ \scriptscriptstyle 0}_2$ can considerably
extend the MSSM parameter space explorable at LEP2.
Although neutralino cross sections are
comparable to chargino-pair production rates only in the
High Cross Section regions, where neutralinos are mostly
higgsinos, the most interesting parameter regions are what
we named Neutralino Regions, where chargino-pair production
is above threshold. In the Neutralino Regions, total rates
for neutralino production crucially depend on selectron
masses. Sizeable rates are obtained mainly in the NR$^+$
for $ \tan \beta $ not too far from 1 and for $m_0
\;\raisebox{-.5ex}{\rlap{$\sim$}} \raisebox{.5ex}{$<$}\;
200$-300 GeV. Depending on the particular scenario selected
in the parameter space, the best channel for neutralino
detection can be either a leptonic or a hadronic one.

Of course, in order to fully assess the potential of
neutralino searches as a tool to discover {\sc SuSy} at LEP2, a
comparative study of the SM processes that can mimic the
neutralino signal has to be performed. This will necessarily
take into account also distributions of relevant kinematical
variables, such as missing momenta and invariant mass of
detected leptonic and hadronic systems.

\acknowledgments

Interesting discussions with Guido Altarelli, Gian
Giudice, Howie Haber, Stavros Katsanevas and Fabio
Zwirner are gratefully acknowledged.

\appendix
\section{}
\label{app:a}

In this Appendix we collect all relevant formulae we use to
calculate sfermion- and Higgs-mass spectrum in the framework
of the MSSM with unification assumptions at the GUT scale.
The neutralino/chargino sector of the model is treated in
Section 2.

For sfermion masses, once the value of $m_0$ is fixed at the
GUT scale, one finds, by performing the RGE evolution down
to the EW scale \cite{ibanez,deboer}:
\begin{equation}
m^2_{\tilde{f}_{L,R}} = \tilde{m}_F^2 + m_f^2 \pm M_D^2,
\label{eq:msf}
\end{equation}
where $m_{\tilde{f}_{L,R}}$ is
the mass of the generic sfermion $\tilde{f}_{L,R}$ and
$\tilde{m}_F$, $m_f$ are the corresponding evolved soft
{\sc SuSy}-breaking mass and fermion mass, respectively. We will
name $\tilde{m}_{Q(L)}$ the soft mass for left squarks
(sleptons) and $\tilde{m}_{U_R \ldots E_R}$ the soft masses
for right squarks and charged leptons. In
\hbox{Eq.~}{}(\ref{eq:msf}), $M_D^2$ is the so-called
``D-term'':
\[ M_D^2 = (T_{3,f_{L,R}} - Q_{f_{L,R}}
\sin^2\!\theta_{ \scriptscriptstyle  W} )  M_{
\scriptscriptstyle  Z} ^2  \cos 2\beta  \; , \]
where
$T_{3,f}$ and $Q_f$ are the SU(2)$_{\rm L}$ and U(1)$_{\rm em}$
(in units of $e > 0$) quantum numbers of the fermion $f$. For
the soft masses of the first two generations,
Yukawa-coupling effects can be neglected and simple formulae
hold. Indeed, they can be expressed, as functions of the
scale $Q$ and in terms of the common scalar and gaugino
masses $m_0$ and $m_{1/2}$ at the GUT scale $M_{\rm
\scriptscriptstyle  GUT}$ (where $\alpha_1 (M_{\rm
\scriptscriptstyle GUT}) = \alpha_2 (M_{\rm
\scriptscriptstyle  GUT}) = \alpha_3 (M_{\rm
\scriptscriptstyle  GUT}) = \alpha_{\rm  \scriptscriptstyle
GUT} \simeq \frac{1}{25}$), through the following equations:
\begin{mathletters}
\label{eq:sfmass}
\begin{eqnarray}
\tilde{m}^2_L(t) & = & m_0^2 + m_{1/2}^2 \frac{\alpha_{\rm
\scriptscriptstyle  GUT}}{4 \pi} \left[ \frac{3}{2} f_2(t) +
\frac{3}{10} f_1(t) \right]       , \label{mL} \\
\tilde{m}^2_{E_R}(t) & = & m_0^2 + m_{1/2}^2
\frac{\alpha_{\rm  \scriptscriptstyle  GUT}}{4 \pi} \left[
\frac{6}{5} f_1(t) \right]                   , \label{mer}
\\ \tilde{m}^2_Q(t) & = & m_0^2 + m_{1/2}^2
\frac{\alpha_{\rm  \scriptscriptstyle  GUT}}{4 \pi} \left[
\frac{8}{3} f_3(t) + \frac{3}{2} f_2(t) + \frac{1}{30}
f_1(t) \right]                                          ,
\label{mQ} \\ \tilde{m}^2_{U_R}(t) & = & m_0^2 + m_{1/2}^2
\frac{\alpha_{\rm  \scriptscriptstyle  GUT}}{4 \pi} \left[
\frac{8}{3} f_3(t) + \frac{8}{15} f_1(t) \right]         ,
\label{mur} \\ \tilde{m}^2_{D_R}(t) & = & m_0^2 + m_{1/2}^2
\frac{\alpha_{\rm  \scriptscriptstyle  GUT}}{4 \pi} \left[
\frac{8}{3} f_3(t) + \frac{2}{15} f_1(t) \right]       ,
\label{mdr}
\end{eqnarray}
\end{mathletters}
where $f_i(t)$
are RGE coefficients at the scale $Q$, given by:
\begin{mathletters}
\label{rge:coeff}
\begin{eqnarray}
f_i(t)  &  =  & \frac{1}{\beta_i} \left( 1 - \frac{1}{(1 +
\beta_i t)^2} \right),  \;\;\;\;  i = 1,2,3,
              \label{rge:f} \\ \beta_i &  =  &
\frac{b_i}{4\pi}\alpha_{\rm  \scriptscriptstyle  GUT},
\;\;\;\;  i = 1,2,3,                 \label{rge:beta} \\ t
    &  = & \log \frac{M^2_{\rm  \scriptscriptstyle
GUT}}{Q^2}. \label{rge:scale}
\end{eqnarray}
\end{mathletters}
In  \hbox{Eq.~}{}(\ref{rge:beta}),
$b_{1,2,3}$ control the evolution of U(1), SU(2), SU(3)
gauge couplings at the one-loop level. Assuming for
simplicity that the whole MSSM particle content contributes
to the evolution from $Q \simeq  M_{ \scriptscriptstyle  Z}
$ up to $M_{\rm  \scriptscriptstyle GUT}$, they are:
\begin{equation}
b_i =         \left( \begin{array}{c} b_1
\\ b_2 \\ b_3 \end{array} \right)     =         \left(
\begin{array}{c} 0 \\ -6 \\ -9 \end{array} \right) + N_{\rm
Fam} \left( \begin{array}{c} 2 \\ 2 \\ 2  \end{array}
\right) + N_{\rm Higgs} \left( \begin{array}{c}3/10 \\ 1/2
\\ 0 \end{array} \right), \label{bmssm}
\end{equation}
where
$N_{\rm Fam} = 3$ is the number of matter supermultiplets
and $N_{\rm Higgs} = 2$ the number of Higgs doublets in the
minimal {\sc SuSy}. Since in the present analysis we use $M_2$ at
the EW scale as an independent parameter in the gaugino
sector, we need also the one-loop RGE relation:
\begin{equation}
M_{1,2,3} ( M_{ \scriptscriptstyle  Z} ) =
\frac{\alpha_{1,2,3} ( M_{ \scriptscriptstyle  Z}
)}{\alpha_{\rm  \scriptscriptstyle  GUT}} m_{1/2}
\Rightarrow  M_3( M_{ \scriptscriptstyle  Z} ) =
\frac{\alpha_3( M_{ \scriptscriptstyle  Z} )}{\alpha_2(
M_{\scriptscriptstyle  Z} )} M_2( M_{ \scriptscriptstyle  Z} )
= \frac{\alpha_3( M_{ \scriptscriptstyle  Z})}
{\alpha_1( M_{ \scriptscriptstyle  Z} )} M_1(
M_{\scriptscriptstyle  Z} ) \; ,
\label{eq:unigauge}
\end{equation}
which allows us to express $m_{1/2}$ in terms
of $M_2$ in  \hbox{Eqs.~}{}(\ref{eq:sfmass}) and from which,
in particular,  \hbox{Eq.~}{}(\ref{eq:unigau}) follows. In
order to properly evaluate the sfermion spectrum through
(\ref{eq:sfmass}), we adopt a recursive procedure (see,
\hbox{e.g.}{},  \hbox{Refs.~}{}\cite{kane93,barger}). First,
for any fixed values of $m_0$ and $M_2$, we calculate {\it
zero-th order} sfermion masses $m_{\tilde{f}}^0$ for $Q= M_{
\scriptscriptstyle  Z} $, then we use these values as an
input in  \hbox{Eqs.~}{}(\ref{eq:sfmass}) (\hbox{i.e.}{},
with $Q=m_{\tilde{f}}^0$ in the corresponding equation for
$\tilde{m}_F$), in order to get out the {\it first order}
masses, and so on. After a few iterations we obtain fast
convergence. In this way, a sufficient agreement with more
sophisticated {\sc SuSy}-spectrum calculations (see,
\hbox{e.g.}{},  \hbox{Ref.~}{}\cite{kane93}) is found. In
all our analysis, we neglect both Yukawa-coupling effects in
diagonal soft masses and left-right mixing for the third
generation of sfermions.

Concerning the {\sc SuSy}-Higgs sector, starting from the two
independent parameters $ m_{A^{ \scriptscriptstyle  0}} $
and $ \tan \beta $, we calculate masses from the relations
\cite{ref:higgs}:
\begin{eqnarray}
(m_{ H^{\scriptscriptstyle  0} , h^{ \scriptscriptstyle  0} })^2
& = & \frac{1}{2} \left[  m_{A^{ \scriptscriptstyle  0}} ^2 +
M_{ \scriptscriptstyle  Z} ^2 +     \Delta \right]  \nonumber \\
& \pm & \sqrt{ \left[ \left(
m_{A^{ \scriptscriptstyle  0}} ^2 - M_{\scriptscriptstyle Z}^2 \right)
\cos 2\beta + \Delta \right]^2  + \left(
m_{A^{ \scriptscriptstyle  0}} ^2 + M_{\scriptscriptstyle Z}^2 \right)^2
\sin^2 2 \beta},  \nonumber \\
m_{H^{\scriptscriptstyle  \pm}} ^2  & = &   m_{A^{\scriptscriptstyle 0}}^2
+  M_{ \scriptscriptstyle  W}^2,
\label{eq:hmass}
\end{eqnarray}
where:
\begin{equation}
\Delta   = \frac{3}{8\pi^2} \frac{g^2 m_t ^4}{ M_{
\scriptscriptstyle  W} ^2 \sin^2\!\beta } \log \left( 1 +
\frac{ m_{\tilde{t}} ^2}{ m_t ^2} \right) .
\eqnum{\ref{eq:hmass}a}
\end{equation}
The  \hbox{Eqs.~}{}(\ref{eq:hmass}) take into account only
the dominant contributions coming from top/s-top loops and
we use it under the further assumptions: $
m_{\tilde{t}_{L,R}}  =  m_{\tilde{u}_{L,R}} $ and no $
\tilde{t}_{ \scriptscriptstyle  L} $-$ \tilde{t}_{
\scriptscriptstyle  R} $ mixing. We found that all the above
simplifications allow us to avoid the introduction of other
{\sc SuSy} parameters as $A_{\rm  \scriptscriptstyle  GUT}$ (or
$A_t( M_{ \scriptscriptstyle  Z} )$) and $B_{\rm
\scriptscriptstyle  GUT}$, without seriously affecting our
results.

\begin{figure}
\caption{Feynman diagrams for the process
$\protect e^{\scriptscriptstyle +}e^{\scriptscriptstyle -}
\rightarrow \protect\tilde{\chi}^{ \scriptscriptstyle  0}_i
\protect\tilde{\chi}^{\scriptscriptstyle  0}_j  $.}
\label{fig:feyprod}
\end{figure}

\begin{figure}
\caption{Photino and Z-ino content in the lightest
neutralino versus $\mu$ and $M_2$, for $ \tan \beta  = 1.5$.}
\label{mix:n1gau}
\end{figure}

\begin{figure}
\caption{Higgsino-A and higgsino-B content in the lightest
neutralino versus $\mu$ and $M_2$, for $ \tan \beta  = 1.5$.}
\label{mix:n1hino}
\end{figure}

\begin{figure}
\caption{Photino and Z-ino content in the next-to-lightest
neutralino versus $\mu$ and $M_2$, for $ \tan \beta  = 1.5$.}
\label{mix:n2gau}
\end{figure}

\begin{figure}
\caption{Higgsino-A and higgsino-B content in the next-to-lightest
neutralino versus $\mu$ and $M_2$, for $ \tan \beta  = 1.5$.}
\label{mix:n2hino}
\end{figure}

\begin{figure}
\caption{Contour plot for the modulus of the LN and NLN mass
eigenvalues (in GeV) in the ($\mu, M_2$) plane for
$ \tan \beta  =1.5$. The dark area corresponds to regions where the
$\protect\tilde{\chi}^{\scriptscriptstyle 0}_{1,2}$-mass eigenvalue
is negative.}
\label{neut12massign}
\end{figure}

\begin{figure}
\caption{Interesting regions and scenarios in the
$(\mu, M_2)$ plane with $ \tan \beta  = 1.5$ for
neutralino search at LEP2 ($\protect\sqrt{s}  = 190$
GeV). The NR$^{\pm}$ regions
(bounded by kinematic-limit curves `N190' and `C190' for
$\protect \tilde{\chi}^{ \scriptscriptstyle  0}_1
\protect\tilde{\chi}^{ \scriptscriptstyle  0}_2 $ and
$\protect\tilde{\chi}^{ \scriptscriptstyle  +}_1
\protect\tilde{\chi}^{\scriptscriptstyle  -}_1 $ production,
respectively) and HCS$^{\pm}$ regions (outlined by the
1 pb contour plot for the
$\protect \tilde{\chi}^{ \scriptscriptstyle  0}_1
\protect \tilde{\chi}^{ \scriptscriptstyle  0}_2 $ total
cross section, for $m_0 =  M_{ \scriptscriptstyle  Z} $) are
indicated. The shaded area corresponds to LEP1 limits. The
points A-D in NR and H$^{\pm}$ in HCS will be used in the
following analysis.}
\label{fig:scentgb15}
\end{figure}

\begin{figure}
\caption{The same as in \protect\hbox{Fig.~}{}
\protect\ref{fig:scentgb15} but with $ \tan \beta  = 4$.
In NR$^{\pm}$ different points are chosen (E, F, G, J) with
respect to the $ \tan \beta  = 1.5$ case.}
\label{fig:scentgb4}
\end{figure}

\begin{figure}
\caption{Contour plot for the total cross section (in fb)
of the process
$\protect e^{ \scriptscriptstyle  +}
\protect e^{ \scriptscriptstyle -}    \rightarrow
\protect\tilde{\chi}^{ \scriptscriptstyle  0}_1
\protect\tilde{\chi}^{ \scriptscriptstyle  0}_2$ at LEP2
($\protect \sqrt{s}  = 190$ GeV) on the ($\mu,M_2$)
plane in the $ \tan \beta  = 1.5$,
$m_0 =  M_{ \scriptscriptstyle  Z}$ case.
The shaded area represents the region excluded by
LEP1 data.}
\label{n1n2prodm91tgb15rs190}
\end{figure}

\begin{figure}
\caption{Contour plot for the total cross section
(in fb) of the process
$\protect e^{\scriptscriptstyle  +}
\protect e^{ \scriptscriptstyle  -}   \rightarrow
\protect\tilde{\chi}^{ \scriptscriptstyle  0}_1
\protect\tilde{\chi}^{ \scriptscriptstyle  0}_2$ at LEP2
($\protect\sqrt{s}  = 190$ GeV) on the ($\mu,M_2$)
plane in the $ \tan \beta  = 1.5$,
$m_0 = 3  M_{ \scriptscriptstyle Z} $ case.}
\label{n1n2prodm273tgb15rs190}
\end{figure}

\begin{figure}
\caption{Contour plot for the total cross section (in fb)
of the process $\protect e^{ \scriptscriptstyle  +}
\protect e^{ \scriptscriptstyle -}    \rightarrow
\protect\tilde{\chi}^{ \scriptscriptstyle  0}_1
\protect\tilde{\chi}^{ \scriptscriptstyle  0}_2$ at LEP2
($\protect\sqrt{s}  = 190$ GeV) on the ($\mu,M_2$)
plane in the $ \tan \beta  = 4$,
$m_0 =  M_{ \scriptscriptstyle  Z} $ case.}
\label{n1n2prodm91tgb4rs190}
\end{figure}

\begin{figure}
\caption{Total cross section (in fb) of the process
$\protect e^{ \scriptscriptstyle  +}
\protect e^{ \scriptscriptstyle -}    \rightarrow
\protect\tilde{\chi}^{ \scriptscriptstyle  0}_1
\protect\tilde{\chi}^{ \scriptscriptstyle  0}_2$ as a
function of $m_0$ (or, once $M_2$ is fixed, of the selectron
masses) in the scenarios defined in
\protect\hbox{Table~}{}\protect\ref{tab:scentgb1p5} for
$ \tan \beta = 1.5$: a) A, B, C, D and b) H$^-$, H$^+$.
Here the \protect\hbox{c.m.}{}\ energy is
$\protect \sqrt{s}  = 190$ GeV.}
\label{prodvsm0tgb1p5}
\end{figure}

\begin{figure}
\caption{Total cross section (in fb) of the process
$\protect e^{ \scriptscriptstyle  +}
\protect e^{ \scriptscriptstyle -}    \rightarrow
\protect\tilde{\chi}^{ \scriptscriptstyle  0}_1
\protect\tilde{\chi}^{ \scriptscriptstyle  0}_2$ as a function
of $\protect \sqrt{s} $ in the LEP2 range in the scenarios
defined in  \protect\hbox{Table~}{}\protect\ref{tab:scentgb1p5},
for $ \tan \beta  = 1.5$: a) A, B, C, D and b) H$^-$, H$^+$.
The selectron masses are fixed by
$m_0 =  M_{ \scriptscriptstyle Z} $. Large symbols are used
when a scenario falls inside the NR$^{\pm}$ regions.}
\label{prodvsrstgb1p5}
\end{figure}

\begin{figure}
\caption{Contour plot for the rate (in fb) of
$\protect e^{\scriptscriptstyle  +}
\protect e^{ \scriptscriptstyle  -} \rightarrow
\protect \tilde{\chi}^{ \scriptscriptstyle  0}_1
\protect \tilde{\chi}^{ \scriptscriptstyle  0}_2
\rightarrow \protect e^{ \scriptscriptstyle  +}
\protect e^{\scriptscriptstyle  -}  +  \not\!\! E $ events
at LEP2 ($\protect \sqrt{s}  = 190$ GeV), in the
case $\tan \beta  =1.5$,
$m_0 =  m_{A^{ \scriptscriptstyle  0}} /3 =
M_{ \scriptscriptstyle  Z} $. Bold curves show
kinematical limits for production of
$\protect\tilde{\chi}^{ \scriptscriptstyle  +}_1
\protect\tilde{\chi}^{ \scriptscriptstyle  -}_1 $
(label `C') and of
$\protect \tilde{\chi}^{ \scriptscriptstyle  0}_1
\protect\tilde{\chi}^{ \scriptscriptstyle  0}_2 $
(label `N'). The shaded area represents the region
excluded by LEP1 data. The thick bold line binds the
region where
$(m_{\protect\tilde{\chi}^0_2} - m_{\protect\tilde{\chi}^0_1} )
< 10$ GeV.}
\label{eveetgb15ma273}
\end{figure}

\begin{figure}
\caption{Contour plot for the rate (in fb) of
$\protect e^{\scriptscriptstyle  +}
\protect e^{ \scriptscriptstyle  -}
\rightarrow \protect \tilde{\chi}^{ \scriptscriptstyle  0}_1
\protect \tilde{\chi}^{ \scriptscriptstyle  0}_2
\rightarrow 2$ jets $+  \not\!\! E $ events.
Notations and parameter values are the same as for
\protect\hbox{Fig.~}{}\protect\ref{eveetgb15ma273}.}
\label{ev2jtgb15ma273}
\end{figure}

\begin{figure}
\caption{Contour plot for the rate (in fb) of
$\protect e^{\scriptscriptstyle  +}
\protect e^{ \scriptscriptstyle  -} \rightarrow
\protect \tilde{\chi}^{ \scriptscriptstyle  0}_1
\protect \tilde{\chi}^{ \scriptscriptstyle  0}_2
\rightarrow \gamma +  \not\!\! E $ events. Notations and
parameter values are the same as for
\protect\hbox{Fig.~}{}\protect\ref{eveetgb15ma273}.}
\label{evphtgb15ma273}
\end{figure}

\begin{figure}
\caption{Contour plot for the total rate (in fb) of visible
events coming from
$\protect \tilde{\chi}^{ \scriptscriptstyle 0}_1
\protect \tilde{\chi}^{ \scriptscriptstyle  0}_2 $
production at LEP2. Notations and parameter values are the
same as for  \protect\hbox{Fig.~}{}\protect\ref{eveetgb15ma273}.}
\label{evvistgb15ma273}
\end{figure}

\begin{figure}
\caption{Contour plot for the rate (in fb) of
$\protect e^{\scriptscriptstyle  +}
\protect e^{ \scriptscriptstyle  -} \rightarrow
\protect \tilde{\chi}^{ \scriptscriptstyle  0}_1
\protect \tilde{\chi}^{ \scriptscriptstyle  0}_2
\rightarrow 2$ jets $+  \not\!\! E $ events at LEP2
($\protect \sqrt{s}  = 190$ GeV), in the case
$\tan \beta  =1.5$,
$m_0 =  m_{A^{ \scriptscriptstyle  0}}  =
M_{ \scriptscriptstyle  Z} $.}
\label{ev2jtgb15ma91}
\end{figure}

\begin{figure}
\caption{Contour plot for the rate (in fb) of
$\protect e^{\scriptscriptstyle  +}e^{ \scriptscriptstyle  -}
\rightarrow \protect \tilde{\chi}^{ \scriptscriptstyle  0}_1
\protect \tilde{\chi}^{ \scriptscriptstyle  0}_2
\rightarrow 4$ jets $+  \not\! E $ at LEP2
($\protect \sqrt{s}  = 190$ GeV), in the case $
\tan \beta  =1.5$, $m_0 =  M_{ \scriptscriptstyle  Z} $ and
$ m_{A^{ \scriptscriptstyle  0}}  =
3  M_{\scriptscriptstyle  Z} $. Inside the bold
dashed line, one has
$(m_{\protect\tilde{\chi}^0_2}  -
m_{\protect\tilde{\chi}^{\pm}_1} ) > 10$ GeV.}
\label{ev4jtgb15ma273}
\end{figure}

\begin{figure}
\caption{Contour plot for the rate (in fb) of
$\protect e^{\scriptscriptstyle  +}
\protect e^{ \scriptscriptstyle  -} \rightarrow
\protect \tilde{\chi}^{ \scriptscriptstyle  0}_1
\protect \tilde{\chi}^{ \scriptscriptstyle  0}_2
\rightarrow \protect e^{ \scriptscriptstyle  +}  + 2$
jets $+  \not\!\! E $ events. Notations and parameter
values are the same as for
\protect\hbox{Fig.~}{}\protect\ref{ev4jtgb15ma273}.}
\label{eve2jtgb15ma273}
\end{figure}

\begin{figure}
\caption{Contour plot for the total rate (in fb) of visible
events coming from
$\protect \tilde{\chi}^{ \scriptscriptstyle 0}_1
\protect \tilde{\chi}^{ \scriptscriptstyle  0}_2 $ at
LEP2 ($\protect \sqrt{s}  = 190$ GeV), in the case
$ \tan \beta = 4$, $m_0 =  m_{A^{ \scriptscriptstyle  0}}
= M_{ \scriptscriptstyle  Z} $.}
\label{evvistgb4ma91}
\end{figure}

\begin{figure}
\caption{Contour plot for the total rate (in fb) of visible
events coming from
$\protect \tilde{\chi}^{ \scriptscriptstyle 0}_1
\protect \tilde{\chi}^{ \scriptscriptstyle  0}_2 $ at
LEP2 ($\protect \sqrt{s}  = 190$ GeV), in the case
$ \tan \beta = 30$, $m_0 =  M_{ \scriptscriptstyle  Z} $, $
m_{A^{ \scriptscriptstyle  0}}  =
3  M_{ \scriptscriptstyle Z} $.}
\label{evvistgb30ma273}
\end{figure}

\begin{figure}
\caption{Contour plot for the rate (in fb) of
$\protect e^{\scriptscriptstyle  +}
\protect e^{ \scriptscriptstyle  -} \rightarrow
\protect \tilde{\chi}^{ \scriptscriptstyle  0}_1
\protect \tilde{\chi}^{ \scriptscriptstyle  0}_2
\rightarrow 4$ jets $+ \not\! E$ at LEP2
($\protect \sqrt{s}  = 190$ GeV), in the case $
\tan \beta  = 30$, $m_0 =  M_{ \scriptscriptstyle  Z} $
and $ m_{A^{ \scriptscriptstyle  0}}  = 3
M_{ \scriptscriptstyle  Z} $.}
\label{ev4jtgb30ma273}
\end{figure}

\begin{figure}
\caption{Contour plot for the total rate (in fb) of visible
events coming from
$\protect \tilde{\chi}^{ \scriptscriptstyle 0}_1
\protect \tilde{\chi}^{ \scriptscriptstyle  0}_2 $
production at the first phase of LEP2
($\protect \sqrt{s}  = 175$ GeV).}
\label{evvisrs175}
\end{figure}

\begin{table}
\caption{Interesting scenarios for neutralino production at
LEP2 ($\protect \sqrt{s}  = 190$ GeV) in the
$ \tan \beta  = 1.5$, $m_0 =  M_{ \scriptscriptstyle  Z} $
case. Mass eigenvalues for charginos and neutralinos are given
as well as sfermion spectrum arising from
$m_0 =  M_{ \scriptscriptstyle  Z} $. For light neutralinos,
the physical composition and the total cross section (in fb)
are also reported for all allowed pair-production processes
$\protect e^{\scriptscriptstyle  +}
\protect e^{ \scriptscriptstyle  -} \rightarrow
\protect \tilde{\chi}^{ \scriptscriptstyle  0}_i
\protect \tilde{\chi}^{ \scriptscriptstyle 0}_j  $.
For the $\protect \tilde{\chi}^{ \scriptscriptstyle  0}_1
\protect \tilde{\chi}^{ \scriptscriptstyle  0}_2 $ case,
individual contributions from $s$-, $(t+u)$-channels and
$(st+su)$ interferences are also given.}

\[ \begin{array}{|c|c||c|c|c|c|c|c|}              \hline
\multicolumn{8}{|c|}{\bf Scenarios  \;\;\;\;  with  \;\;\;\;
  \tan \beta  = 1.5} \\ \hline\hline
\multicolumn{2}{|c||}{\bf Scenario} & {\bf A} & {\bf B} &
{\bf C} & {\bf D} & {\bf H^-} & {\bf H^+}  \\ \hline
\multicolumn{2}{|c||}{(\mu,\: M_2)/ M_{ \scriptscriptstyle
Z}  \;\;   \rightarrow  } & (-3, \: 1) & (-1, \: 1) & (-1,
\: 1.5) & (3, \: 1.5) & (-0.7, \: 3) & (1, \: 3)   \\ \hline
\multicolumn{2}{|c||}{M_1 \; {\rm (GeV)}  \;\;\;\;  \rightarrow }
&  45.7 &  45.7 &  68.6 & 68.6 & 137.2 & 137.2       \\ \hline\hline
&  {\rm Mass \: (GeV)} &  49.5 &  51.5 &  73.7 &   56.0 &  62.3 &  44.9
  \\ \cline{2-8}
\tilde{\chi}^{ \scriptscriptstyle  0}_1
        & ( \tilde{\gamma} ,\:  \tilde{Z} )  \;\;\;  (\%)
        & (88,\: 11) & (91,\: 6) & (76,\: 10) & (47,\: 45)
& ( 0,\:  1)          & ( 4,\: 20) \\ \cline{2-8}
& ( \tilde{H}^{ \scriptscriptstyle  0}_a ,\:
\tilde{H}^{ \scriptscriptstyle  0}_b ) \; (\%)
& ( 1,\:  0) &  (2,\: 2) & ( 1,\: 13) & ( 7,\:  1) & ( 2,\: 97)
& (70, \: 5)                            \\ \hline
& {\rm Mass \: (GeV)}  & 107.0 & 85.2 &  89.8 &  108.2 & -89.1 & -92.3
                                        \\ \cline{2-8}
\tilde{\chi}^{ \scriptscriptstyle  0}_2
& (\tilde{\gamma} ,\:  \tilde{Z} )  \;\;\;  (\%)
& (12,\: 83) & ( 4,\: 9) & (15,\:  1) & (53,\: 36) & ( 0,\: 7)
& ( 0,\:  0)     \\ \cline{2-8}
& ( \tilde{H}^{\scriptscriptstyle  0}_a ,\:
\tilde{H}^{ \scriptscriptstyle 0}_b ) \; (\%)
& ( 4,\:  2) & ( 0,\: 86) & ( 2,\:
83) & (10,\:  1) & (90,\: 2)    & ( 5,\: 94) \\ \hline
\multicolumn{2}{|c||}{ \tilde{\chi}^{ \scriptscriptstyle 0}_3
\;\;\;\;\;\;\;\;\;\;  \; {\rm Mass \: (GeV)}}
& 275.4 & -129.8 & -124.5 & -274.4 & 144.9 & 153.5 \\ \hline
\multicolumn{2}{|c||}{ \tilde{\chi}^{\scriptscriptstyle  0}_4
\;\;\;\;\;\;\;\;\;\;  \; {\rm Mass \: (GeV)}}
& -294.9 &  130.0 & 166.4 &  315.6 & 292.6 & 304.6   \\ \hline
\multicolumn{2}{|c||}{ \tilde{\chi}^{ \scriptscriptstyle  \pm}_1
\;\;\;\;\;\;\;\;\;\;  \; {\rm Mass \: (GeV)}} &  106.1 &
104.7 & 110.8 & -101.5 &  80.1 & -62.6             \\ \hline
\multicolumn{2}{|c||}{ \tilde{\chi}^{ \scriptscriptstyle \pm}_2
\;\;\;\;\;\;\;\;\;\;  \; {\rm Mass \: (GeV)}}
& 291.2 &  136.2 & 166.2 &  310.0 & 292.2 & 303.5  \\ \hline\hline
\multicolumn{2}{|c||}{ \tilde{e}_{\scriptscriptstyle  L} ,
\tilde{e}_{ \scriptscriptstyle  R},
\tilde{\nu}_{ \scriptscriptstyle  e,L}
\;\;\;\;\;\;  {\rm Mass \: (GeV)}  \;\; } &
\multicolumn{2}{c|}{124, 104, 114} &
\multicolumn{2}{c|}{152,115,144} &
\multicolumn{2}{c|}{255,160,250}                        \\ \hline
\multicolumn{2}{|c||}{ \tilde{u}_{ \scriptscriptstyle  L} ,
\tilde{u}_{ \scriptscriptstyle  R}  \;\;\;\;\;\;\;\;
{\rm Mass \: (GeV)}} & \multicolumn{2}{c|}{285, 277} &
\multicolumn{2}{c|}{408,395} & \multicolumn{2}{c|}{773,746} \\ \hline
\multicolumn{2}{|c||}{ \tilde{d}_{\scriptscriptstyle  L} ,
\tilde{d}_{ \scriptscriptstyle  R}
 \;\;\;\;\;\;\;\;  {\rm Mass \: (GeV)}} &
\multicolumn{2}{c|}{289,278} & \multicolumn{2}{c|}{411,395}
& \multicolumn{2}{c|}{774,743}              \\ \hline\hline
\multicolumn{2}{|c||}{\sigma( \tilde{\chi}^{\scriptscriptstyle  0}_1
\tilde{\chi}^{ \scriptscriptstyle 0}_1 )
\;\;\;\;\;\;\;\;\;\;  {\rm Total \: (fb)}}
 & 881.7 & 770.9 & 213.2 &  448.3 &   6.6 &   6.7 \\\hline
 & {\rm Total}            & 146.4 & 112.8 &
 80.9 &   56.1 &  1654 &  1400 \\ \cline{2-8}
\sigma(\tilde{\chi}^{ \scriptscriptstyle  0}_1
\tilde{\chi}^{ \scriptscriptstyle  0}_2 ) & {\rm (s)-channel}
& 0.1 &  18.1 &   9.1 &   0.1 &  1622 &  1366     \\ \cline{2-8}
{\rm (fb)} & {\rm (t+u)-channels}            &
142.7 & 48.1 &  38.9 &   51.8 &   0.3 &   0.4     \\ \cline{2-8}
 & {\rm Interferences}
&   3.6 & 46.5 &  32.8 & 4.2 &  31.0 & 33.4     \\ \hline
\multicolumn{2}{|c||}{\sigma( \tilde{\chi}^{\scriptscriptstyle  0}_1
\tilde{\chi}^{ \scriptscriptstyle 0}_3 )
 \;\;\;\;\;\;\;\;\;\;  {\rm Total \: (fb)}}
 &   --  & 176.7 &   --  &    --  &   --  &   --     \\ \hline
\multicolumn{2}{|c||}{\sigma( \tilde{\chi}^{\scriptscriptstyle  0}_1
\tilde{\chi}^{ \scriptscriptstyle 0}_4 )
 \;\;\;\;\;\;\;\;\;\;  {\rm Total \: (fb)}}
 &   --  &  22.3 &   --  &    --  &   --  &   --   \\ \hline
\multicolumn{2}{|c||}{\sigma( \tilde{\chi}^{\scriptscriptstyle  0}_2
\tilde{\chi}^{ \scriptscriptstyle 0}_2 )
\;\;\;\;\;\;\;\;\;\;  {\rm Total \: (fb)}}
 &   --  &   4.8 &  0.4  &    --  &   0.3 &   0.1    \\ \hline
\end{array} \]
\label{tab:scentgb1p5}
\end{table}

\begin{table}
\caption{Interesting scenarios for neutralino production at
LEP2 ($\protect \sqrt{s}  = 190$ GeV) in the
$ \tan \beta  = 4$, $m_0 =  M_{ \scriptscriptstyle  Z} $ case.
See \protect\hbox{Table~}{}\protect\ref{tab:scentgb1p5} for
explanations.}

\[ \begin{array}{|c|c||c|c|c|c|c|c|}              \hline
\multicolumn{8}{|c|}{\bf Scenarios  \;\;\;\;  with  \;\;\;\;
  \tan \beta  = 4}   \\ \hline\hline
\multicolumn{2}{|c||}{\bf Scenario} & {\bf E} & {\bf F} &
{\bf G} & {\bf J} & {\bf H^-} & {\bf H^+}  \\ \hline
\multicolumn{2}{|c||}{(\mu,\: M_2)/ M_{ \scriptscriptstyle
Z}  \;\;   \rightarrow  } & (-3, \: 1.1) & (-1.5, \: 1.5) &
(2, \: 1.7) & (3, \: 1.3) & (-0.7, \: 3) & (1, \: 3)
               				\\ \hline
\multicolumn{2}{|c||}{M_1 \; {\rm (GeV)}  \;\;\;\;
\rightarrow }          &  50.3 &  68.6 &  77.7 & 59.4 &
137.2 & 137.2       \\ \hline\hline
&  {\rm Mass \: (GeV)}
&  52.1 &  68.1 &  62.7 &   53.0 &  53.5 &  55.7    \\ \cline{2-8}
\tilde{\chi}^{ \scriptscriptstyle  0}_1  &
( \tilde{\gamma} ,\:  \tilde{Z} )  \;\;\; (\%)          &
(81,\: 17) & (66,\: 21) & (40,\: 41) & (57,\: 37) & ( 1,\:
5)          & ( 4,\: 16)                            \\ \cline{2-8}
& ( \tilde{H}^{\scriptscriptstyle  0}_a ,\:
\tilde{H}^{ \scriptscriptstyle 0}_b ) \; (\%)
& ( 0,\:  2) &  (0,\: 13) & ( 9,\: 10) & ( 2,\:  3) & (14,\: 80)
& (48,\: 32)                                          \\ \hline
& {\rm Mass \: (GeV)} & 102.9 & 107.3 & 113.8 &   99.4 & -84.3 & -98.6
                                                     \\ \cline{2-8}
\tilde{\chi}^{\scriptscriptstyle  0}_2  & ( \tilde{\gamma} ,\:
\tilde{Z})  \;\;\;  (\%)
& (19,\: 72) & (30,\: 19) & (58,\: 18) & (43,\: 47) & ( 0,\: 6)
& ( 0,\:  2)                                          \\ \cline{2-8}
& ( \tilde{H}^{ \scriptscriptstyle 0}_a ,\:
\tilde{H}^{ \scriptscriptstyle  0}_b ) \; (\%)
& ( 0,\:  8) & ( 6,\: 45) & (15,\:  9) & ( 5,\:  5)
& (75,\:18)          & (32,\: 66)   \\ \hline
\multicolumn{2}{|c||}{ \tilde{\chi}^{ \scriptscriptstyle
0}_3   \;\;\;\;\;\;\;\;\;\; {\rm Mass \: (GeV)}}
&  285.5 & -159.5 & -189.6 & -279.5 & 146.8 & 152.1  \\ \hline
\multicolumn{2}{|c||}{ \tilde{\chi}^{ \scriptscriptstyle  0}_4
\;\;\;\;\;\;\;\;\;\;  {\rm Mass \: (GeV)}}
& -289.9 &  189.4 &  245.8 & 305.1 & 294.8 & 301.6  \\ \hline
\multicolumn{2}{|c||}{ \tilde{\chi}^{ \scriptscriptstyle  \pm}_1
\;\;\;\;\;\;\;\;\;\;  {\rm Mass \: (GeV)}}            &
103.2 &  111.8 & -103.5 & -96.7 &  69.5 & -72.8  \\ \hline
\multicolumn{2}{|c||}{ \tilde{\chi}^{ \scriptscriptstyle \pm}_2
\;\;\;\;\;\;\;\;\;\;  {\rm Mass \: (GeV)}}
&  295.2 &  194.4 &  243.8 & 304.0 & 294.9 & 301.2  \\ \hline\hline
\multicolumn{2}{|c||}{ \tilde{e}_{ \scriptscriptstyle  L}
\;\;\;\;\;\;\;\;\;\;  {\rm Mass \: (GeV)}}
&  134  &  156  &  168   &  144   & \multicolumn{2}{c|}{257} \\ \hline
\multicolumn{2}{|c||}{ \tilde{e}_{ \scriptscriptstyle  R}
\;\;\;\;\;\;\;\;\;\;  {\rm Mass \: (GeV)}}
&  111  &  119  &  124   &  115   & \multicolumn{2}{c|}{163} \\ \hline
\multicolumn{2}{|c||}{ \tilde{\nu}_{ \scriptscriptstyle  e,L}
\;\;\;\;\;\;\;\;  {\rm Mass \: (GeV)}}
&  111  &  137  &  151   &  124   & \multicolumn{2}{c|}{246} \\ \hline
\multicolumn{2}{|c||}{ \tilde{u}_{ \scriptscriptstyle  L}
\;\;\;\;\;\;\;\;\;\;  {\rm Mass \: (GeV)}}
&  307  &  406  &  455   &  357   & \multicolumn{2}{c|}{772} \\ \hline
\multicolumn{2}{|c||}{ \tilde{u}_{ \scriptscriptstyle  R}
\;\;\;\;\;\;\;\;\;\;  {\rm Mass \: (GeV)}}
&  300  &  395  &  442   &  347   & \multicolumn{2}{c|}{745} \\ \hline
\multicolumn{2}{|c||}{ \tilde{d}_{ \scriptscriptstyle  L}
\;\;\;\;\;\;\;\;\;\;  {\rm Mass \: (GeV)}}
&  316  &  413  &  461   &  364   & \multicolumn{2}{c|}{775} \\ \hline
\multicolumn{2}{|c||}{ \tilde{d}_{ \scriptscriptstyle  R}
\;\;\;\;\;\;\;\;\;\;  {\rm Mass \: (GeV)}}
&  301  &  395  &  442   &  348   & \multicolumn{2}{c|}{743} \\ \hline\hline
\multicolumn{2}{|c||}{\sigma( \tilde{\chi}^{ \scriptscriptstyle  0}_1
\tilde{\chi}^{ \scriptscriptstyle  0}_1 )  \;\;\;\;\;\;\;\;\;\;
{\rm Total \: (fb)}}
& 731.7 & 268.3 & 215.9 &  545.3 & 42.7 &  19.8     \\ \hline
& {\rm Total}  & 111.2 &  40.6 &  33.8 &  88.7 &  1688 &  1307  \\ \cline{2-8}
\sigma( \tilde{\chi}^{ \scriptscriptstyle  0}_1
\tilde{\chi}^{ \scriptscriptstyle  0}_2 ) & {\rm (s)-channel}
&   0.7 &   3.7 &   0.9 &  1.0 &  1617 &  1236      \\ \cline{2-8}
{\rm (fb)} & {\rm (t+u)-chans.}
&  99.3 & 20.9 &  25.4 &  71.6 &   1.6 &   2.1      \\ \cline{2-8}
& {\rm Interferences}
&  11.1 &  16.1 &   7.5 &  16.1 &  69.7 & 69.1      \\ \hline
\multicolumn{2}{|c||}{\sigma( \tilde{\chi}^{ \scriptscriptstyle  0}_2
\tilde{\chi}^{ \scriptscriptstyle  0}_2 )  \;\;\;\;\;\;\;\;\;\;
{\rm Total \: (fb)}}
&   --  & --  &   --  &   --  &   3.3 &    --      \\ \hline \end{array} \]
\label{tab:scentgb4}
\end{table}

\begin{table}
\caption{Branching Ratios (\%) for
$\protect \tilde{\chi}^{ \scriptscriptstyle  0}_2 $ decays
in the scenarios with $ \tan \beta  = 1.5$, defined in Section 3.
Sfermion masses are fixed by $m_0 =  M_{ \scriptscriptstyle  Z} $
and the indicated value of $ m_{A^{ \scriptscriptstyle  0}} $ sets
the Higgs spectrum.}

\[ \begin{array}{|c||c|c||c|c||c|c||c||c||c|c|} \hline
\multicolumn{11}{|c|}{\bf Branching  \;\; Ratios  \;\;  (\%)
 \;\;  for  \;\;   \tilde{\chi}^{ \scriptscriptstyle  0}_2
\;\; decays  \;\;  ( \tan \beta  = 1.5)} \\ \hline \hline
{\bf Scenario} & \multicolumn{2}{c||}{\bf A} &
\multicolumn{2}{c||}{\bf B} & \multicolumn{2}{c||}{\bf C} &
{\bf D} & {\bf H^-} & \multicolumn{2}{c|}{\bf H^+}
 \\ \hline
(\mu,\: M_2)/ M_{ \scriptscriptstyle  Z}
\rightarrow  & \multicolumn{2}{c||}{(-3, \: 1)} &
\multicolumn{2}{c||}{(-1, \: 1)} & \multicolumn{2}{c||}{(-1,
\: 1.5)} & (3, \: 1.5) & (-0.7, \: 3) &
\multicolumn{2}{c|}{(1, \: 3)} \\ \hline
m_{A^{\scriptscriptstyle  0}}  \: {\rm (GeV)} \;  \rightarrow  &
M_{ \scriptscriptstyle  Z}  & 3 M_{ \scriptscriptstyle  Z} &
 M_{ \scriptscriptstyle  Z}  & 3 M_{ \scriptscriptstyle  Z}
&  M_{ \scriptscriptstyle  Z}  & 3 M_{ \scriptscriptstyle
Z}  &  \;\raisebox{-.5ex}{\rlap{$\sim$}}
\raisebox{.5ex}{$>$}\;  75 & \;\raisebox{-.5ex}{\rlap{$\sim$}}
\raisebox{.5ex}{$>$}\;  27 &  M_{ \scriptscriptstyle  Z} /2 &
M_{ \scriptscriptstyle Z} \\ \hline \hline
\tilde{\chi}^{ \scriptscriptstyle 0}_1
h^{ \scriptscriptstyle  0}  & 97.6 &  --  &   -- &
-- &   -- &  --  &  --  &  --  &  --  & -- \\ \hline
\tilde{\chi}^{ \scriptscriptstyle  0}_1
A^{ \scriptscriptstyle  0}  &  --  &  --  &   -- & -- &   -- &
--  &  --  &  --  & 21.2 &  -- \\ \hline
\tilde{\chi}^{\scriptscriptstyle  0}_1  e^{ \scriptscriptstyle  +}
e^{\scriptscriptstyle  -}  &  0.6 & 26.6 &  2.8 &  2.7 &  9.9 &
 9.7 & 25.9 &  3.3 &  2.0 & 2.5 \\ \hline
\sum_{\ell} \tilde{\chi}^{ \scriptscriptstyle  0}_1
\nu_{ \scriptscriptstyle  \ell}
\bar{\nu}_{ \scriptscriptstyle \ell}
&  0.5 & 19.4 &  6.9 & 6.8 & 12.6 & 12.3 & 20.0 &
20.3 & 12.1 & 15.3 \\ \hline
\sum_{q \ne t} \tilde{\chi}^{ \scriptscriptstyle  0}_1
q\bar{q}  &  --  &  0.5 & 80.9 & 79.8 & 51.7 & 50.6
&  1.7 & 67.7 & 40.2 & 51.0 \\ \hline
\tilde{\chi}^{ \scriptscriptstyle  \pm}_1
e^{ \scriptscriptstyle  \mp}  \bar{\nu}_{ \scriptscriptstyle  e}
( \nu_{ \scriptscriptstyle  e} ) &  --  & --  &  --  &  --
&  --  &  --  &  0.1 &  0.2 & 2.3 &  2.9 \\ \hline
\sum_{(q,q^{\prime}) \ne t}  \tilde{\chi}^{\scriptscriptstyle \pm}_1
q \bar{q}^{\prime} &  --  &  --  &  -- &  --  &  --
&  --  &  0.2 &  1.1 & 13.7 & 17.4 \\ \hline
\tilde{\chi}^{ \scriptscriptstyle  0}_1  \gamma &
--  &  0.2 &  3.8 & 5.1 &  6.0 &  7.9 &  0.1 &  0.5 &  --  &
 --  \\ \hline
{\rm All \; visible} & 99.5 & 80.6 & 93.1 &
93.2 & 87.4 & 87.7 & 80.0 & 79.7 & 87.9 & 84.7 \\ \hline
\end{array} \]
\label{tab:brtgb1p5}
\end{table}

\begin{table}
\caption{The same as in
\protect\hbox{Table~}{}\protect\ref{tab:brtgb1p5},
but for $ \tan \beta  = 4$.}

\[ \begin{array}{|c||c|c||c||c|c||c|c||c||c|} \hline
\multicolumn{10}{|c|}{\bf Branching  \;\; Ratios  \;\;  (\%)
 \;\;  for  \;\;   \tilde{\chi}^{ \scriptscriptstyle  0}_2
\;\; decays  \;\;  ( \tan \beta  = 4)} \\ \hline \hline
{\bf Scenario} & \multicolumn{2}{c||}{\bf E} & {\bf F} &
\multicolumn{2}{c||}{\bf G} & \multicolumn{2}{c||}{\bf J} &
{\bf H^-} & {\bf H^+} 	\\ \hline
(\mu,\: M_2)/ M_{ \scriptscriptstyle  Z}  \rightarrow
& \multicolumn{2}{c||}{(-3, \: 1.1)} & (-1.5, \: 1.5)
& \multicolumn{2}{c||}{(2, \: 1.7)}
& \multicolumn{2}{c||}{(3, \: 1.3)} & (-0.7, \: 3) & (1, \: 3)
\\ \hline
m_{A^{ \scriptscriptstyle  0}}  \: {\rm (GeV)} \; \rightarrow
&  M_{ \scriptscriptstyle  Z} /2 &  M_{ \scriptscriptstyle  Z}
& {\rm any} &  M_{ \scriptscriptstyle Z} /2
&  M_{ \scriptscriptstyle  Z}
&  M_{ \scriptscriptstyle  Z} /2  &  M_{ \scriptscriptstyle  Z}
& {\rm any}  &  \;\raisebox{-.5ex}{\rlap{$\sim$}}
\raisebox{.5ex}{$>$}\;  43 \\ \hline \hline
\tilde{\chi}^{ \scriptscriptstyle  0}_1   h^{\scriptscriptstyle  0}
& 97.9 &   --  &   -- & 98.6 &  --  & 99.8 &  --  &  --  &  --
\\ \hline
\tilde{\chi}^{ \scriptscriptstyle  0}_1   A^{ \scriptscriptstyle  0}
&  1.8 &   --  &   --  &  1.4 &  --  &  0.2 &  --  &  --  &  --
\\ \hline
\tilde{\chi}^{ \scriptscriptstyle  0}_1  e^{ \scriptscriptstyle  +}
e^{ \scriptscriptstyle  -}  &  --  & 8.8  & 11.1  &  --
& 25.6 &  --  & 12.3 &  3.0  &  2.6  \\ \hline
\sum_{\ell} \tilde{\chi}^{ \scriptscriptstyle  0}_1
\nu_{ \scriptscriptstyle  \ell}  \bar{\nu}_{ \scriptscriptstyle  \ell}
&  0.2 & 70.1  &  0.6 &  --  &  1.4 &  --  & 55.6 & 18.9  & 16.1
\\ \hline
\sum_{q \ne t} \tilde{\chi}^{ \scriptscriptstyle  0}_1   q\bar{q}
&  --  & 3.4  & 65.6  &  --  & 19.1 &  --  &  7.4 & 62.6  & 53.1 \\ \hline
\tilde{\chi}^{ \scriptscriptstyle  \pm}_1  e^{ \scriptscriptstyle  \mp}
\bar{\nu}_{ \scriptscriptstyle  e} ( \nu_{ \scriptscriptstyle  e} )
&  --  &  --   &  --   &  --  &  0.4 &  --  &  --  &  1.0  &  2.6 \\ \hline
\sum_{(q,q^{\prime}) \ne t}  \tilde{\chi}^{ \scriptscriptstyle  \pm}_1
q \bar{q}^{\prime} &  --  &  --  &  --   &  --  & 1.7
&  --  &  --  &  6.1  & 15.4  \\ \hline
\tilde{\chi}^{ \scriptscriptstyle  0}_1  \gamma & --  &  0.2
&  0.6  &  --  & --  &  --  &  --  &  0.2  &  --   \\ \hline
{\rm All \; visible} & 99.8 & 29.9  & 99.4  & 100 & 98.6 &  100
& 44.4 & 81.1  & 83.9  \\ \hline
\end{array} \]
\label{tab:brtgb4}
\end{table}

\begin{table}
\caption{Total rates (in fb) corresponding to different
signatures arising from
$\protect \tilde{\chi}^{ \scriptscriptstyle  0}_1
\protect \tilde{\chi}^{ \scriptscriptstyle  0}_2 $
production at LEP2 ($\protect \sqrt{s}  = 190$ GeV)
in the six significant scenarios with $ \tan \beta  = 1.5$
defined in Section 3.
Sfermion masses are fixed by
$m_0 =  M_{ \scriptscriptstyle Z}$ and the indicated value of
$ m_{A^{ \scriptscriptstyle 0}} $ sets the Higgs spectrum.}

\[ \begin{array}{|c||c|c||c|c||c|c||c||c||c|c|} \hline
\multicolumn{11}{|c|}{\bf Rates  \;\;  (fb)  \;\;  for  \;\;
\protect e^{ \scriptscriptstyle  +}
\protect e^{ \scriptscriptstyle -}  \rightarrow
\protect \tilde{\chi}^{ \scriptscriptstyle 0}_1
\protect \tilde{\chi}^{ \scriptscriptstyle  0}_2
\rightarrow   \;\;  final  \;\;  state  \;\;
( \tan \beta = 1.5)} \\ \hline \hline
{\bf Scenario} & \multicolumn{2}{c||}{\bf A} & \multicolumn{2}{c||}{\bf B}
& \multicolumn{2}{c||}{\bf C} & {\bf D} & {\bf H^-} &
\multicolumn{2}{c|}{\bf H^+}            \\ \hline
(\mu,\: M_2)/ M_{ \scriptscriptstyle  Z}   \rightarrow
& \multicolumn{2}{c||}{(-3, \: 1)}
& \multicolumn{2}{c||}{(-1,\: 1)}
& \multicolumn{2}{c||}{(-1, \: 1.5)} & (3, \: 1.5) & (-0.7, \: 3)
& \multicolumn{2}{c|}{(1, \: 3)} \\ \hline
m_{A^{ \scriptscriptstyle  0}}  \: {\rm (GeV)} \;
\rightarrow  &  M_{ \scriptscriptstyle  Z}
& 3 M_{ \scriptscriptstyle  Z}   &  M_{ \scriptscriptstyle  Z}
& 3 M_{ \scriptscriptstyle  Z}   &  M_{ \scriptscriptstyle  Z}
& 3 M_{ \scriptscriptstyle  Z}   &  \;\raisebox{-.5ex}{\rlap{$\sim$}}
\raisebox{.5ex}{$>$}\;  75 &  \;\raisebox{-.5ex}{\rlap{$\sim$}}
\raisebox{.5ex}{$>$}\; 27 &  M_{ \scriptscriptstyle  Z} /2
& M_{ \scriptscriptstyle  Z} \\ \hline \hline
e^{ \scriptscriptstyle  +}e^{ \scriptscriptstyle  -}  +  \not\!\! E
&   0.9 & 39.0 &  3.1 &   3.1 &   8.0 &   7.8 &  14.6 & 55.5 &  31.3
&  39.7 \\ \hline
e^{ \scriptscriptstyle  +}  \mu^{ \scriptscriptstyle  -}  +  \not\!\! E
&   --  &   --  &   -- &    -- &    -- &   --  &  --  &   0.3 &   3.6
&   4.6 \\ \hline
{\rm Invisible} &   0.7 &  28.4 &   7.8 &   7.7 & 10.2 &  10.0 &  11.2 & 335.2
& 168.9 & 214.2 \\ \hline
\tau^{ \scriptscriptstyle  +}\tau^{ \scriptscriptstyle  -}
+  \not\!\! E  &   6.9 &  39.0 &   3.1 &   3.1 &   8.0 &
7.8 &  14.6 &  55.5 &  31.9 &  39.7 \\ \hline
2 j\: +  \not\!\! E  & 136.9 &   0.7 &  91.2 & 90.0 &  41.8 &  40.9 &
  1.0 &  1120 & 858.9 & 714.2 \\ \hline  b\bar{b} \: +
\not\!\! E  & 133.5 &   --  &  18.5 &  18.2 &   8.9 &   8.7
&   0.3 & 246.0 & 419.3 & 156.9 \\ \hline
e^{ \scriptscriptstyle  +}  + 2j\: +  \not\!\! E  & --  &   --
&   --  &   --  &   --  &   --  & -- &   2.0 &  21.5 &  27.2
\\ \hline
4 j\: +  \not\!\! E  &   --  &   --  &   --  &  --  &   --  &   --
&   0.1 &  11.6 & 127.5 & 161.7 \\ \hline
\gamma\: +  \not\!\! E  &   --  &   0.3 &   4.3 & 5.8 &   4.8 &   6.4
&   --  &   6.5 &   --  &   -- \\ \hline
{\rm All \; visible} & 145.7 & 118.0 & 104.9 & 105.0 &  70.7
&  70.9 &  44.9 &  1318 & 1231 & 1186 \\ \hline
\end{array} \]
\label{tab:evtgb1p5}
\end{table}

\begin{table}
\caption{The same as in
\protect\hbox{Table~}{}\protect\ref{tab:evtgb1p5},
but for $ \tan \beta  = 4$.}

\[ \begin{array}{|c||c|c||c||c|c||c|c||c||c|} \hline
\multicolumn{10}{|c|}{\bf Rates  \;\;  (fb)  \;\;  for  \;\;
\protect e^{ \scriptscriptstyle  +}
\protect e^{ \scriptscriptstyle -}  \rightarrow
\protect \tilde{\chi}^{ \scriptscriptstyle 0}_1
\protect \tilde{\chi}^{ \scriptscriptstyle  0}_2
\rightarrow   \;\;  final  \;\;  state  \;\;
( \tan \beta  = 4)} \\ \hline \hline
{\bf Scenario} & \multicolumn{2}{c||}{\bf E} & {\bf F}
& \multicolumn{2}{c||}{\bf G} & \multicolumn{2}{c||}{\bf J}
& {\bf H^-} & {\bf H^+}         \\ \hline
(\mu,\: M_2)/ M_{ \scriptscriptstyle  Z}   \rightarrow
& \multicolumn{2}{c||}{(-3, \: 1.1)} & (-1.5, \: 1.5)
& \multicolumn{2}{c||}{(2, \: 1.7)}
& \multicolumn{2}{c||}{(3, \: 1.3)} & (-0.7, \: 3) & (1, \: 3)
\\ \hline
m_{A^{ \scriptscriptstyle  0}}  \: {\rm (GeV)} \;  \rightarrow
&  M_{ \scriptscriptstyle  Z} /2   &  M_{ \scriptscriptstyle  Z}
& {\rm any} &  M_{ \scriptscriptstyle  Z} /2
&  M_{ \scriptscriptstyle  Z}  &  M_{ \scriptscriptstyle  Z} /2
& M_{ \scriptscriptstyle  Z}  & {\rm any}
& \;\raisebox{-.5ex}{\rlap{$\sim$}} \raisebox{.5ex}{$>$}\;  43
\\ \hline \hline
e^{ \scriptscriptstyle  +} e^{ \scriptscriptstyle  -}  +
\not\!\! E  &   --  &   9.7 &   4.5 &   --  &   8.6 &   --
&  10.9 &  53.4 &  37.3 \\ \hline
e^{ \scriptscriptstyle +}  \mu^{ \scriptscriptstyle  -}  +  \not\!\! E
&   --  &  --  &    -- &   --  &    -- &   --  &   --  &   1.9 &   3.8
\\ \hline
{\rm Invisible} &   0.2 &  77.9 &   0.3 &   --  &  0.5 &   --
&  49.3 & 319.6 & 210.5 \\ \hline
\tau^{ \scriptscriptstyle  +}\tau^{ \scriptscriptstyle  -}  +
\not\!\! E  &   4.8 &   9.7 &   4.5 &   1.4 &   8.6 &   3.9
&  10.9 &  53.4 &  37.3 \\ \hline
2 j\: +  \not\!\! E  & 106.1 &   3.8 &  26.6 &  32.3 &   6.4 &  84.8
&   6.6 & 1057 & 693.4 \\ \hline  b\bar{b} \: +  \not\!\! E  & 105.9 &
  0.7  &  5.8 &  32.2 &   1.5 &  84.6 &   1.6 & 232.0 &
152.2 \\ \hline
e^{ \scriptscriptstyle  +}  + 2j\: +  \not\!\! E
&   --  &   --  &   --  &   --  &   0.1 &   --
&   --  &  11.5 &  22.5 \\ \hline
4 j\: +  \not\!\! E  &  --  &   --  &   --  &   --  &   0.3
&   --  &   --  &  68.3 & 133.3 \\ \hline
\gamma\: +  \not\!\! E  &   --  &   0.2 &  0.2 &   --  &   --
&   --  &   --  &   2.8 &   0.1 \\ \hline
{\rm All \; visible} & 110.9 &  33.2 &  40.4 &  33.7 & 33.3
&  88.7 &  39.4 &  1369 &  1097 \\ \hline
\end{array}\]
\label{tab:evtgb4}
\end{table}


\begin{references}

\bibitem{kane93} G.~L.~Kane, C.~Kolda, L.~Roszkowski and J.~D.~Wells,
        Phys. Rev. {\bf D49}, 6173 (1994); {\bf D50}, 3498 (1994).

\bibitem{reviews} For a review and references see:
        H.-P.~Nilles, Phys. Rep. {\bf 110}, 1 (1984);
        R.~Barbieri, Riv. Nuovo Cim. {\bf 11}, 1 (1988).

\bibitem{hk} H.~E.~Haber and G.~L.~Kane,
        Phys. Rep. {\bf 117}, 75 (1985).

\bibitem{lep1neut} R.~Barbieri, G.~Gamberini, G.~F.~Giudice
        and G.~Ridolfi, Phys. Lett. {\bf 195B}, 500 (1987);
        J.~Ellis, G.~Ridolfi and F.~Zwirner,
        {\it ibid.} {\bf 237B}, 423 (1990).

\bibitem{bartl86a} A.~Bartl, H.~Fraas, W.~Majerotto,
        Nucl. Phys. {\bf B278}, 1 (1986);
        A.~Bartl, W.~Majerotto, B.~M\"osslacher, Proc. of the
        Joint Int. Lepton-Photon Symp. \& Europhys. Conf. on
        High Energy Phys., Geneva, Jul. 25-Aug. 1, 1991, vol. I, p.357.

\bibitem{dionisi} D.~A.~Dicus and X.~Tata,
        Phys. Rev. {\bf D35}, 2110 (1987);
	M.~Chen, C.~Dionisi, M.~Martinez, X.~Tata,
        Phys. Rep. {\bf 159}, 201 (1988);
	J.~F.~Grivaz, C.~Dionisi  \hbox{\it et al.}{}, in Proc. of the
        ECFA Workshop on LEP200, Aachen, Sept. 29-Oct. 1, 1986,
        A.~B\"ohm and W.~Hoogland eds., CERN 87-08, ECFA 87/108, 1987,
        Vol. II, p.380.

\bibitem{ambmele} S.~Ambrosanio and B.~Mele, Preprint ROME1-1095/95.

\bibitem{neumatr} J.~Ellis and G.~G.~Ross,
        Phys. Lett. {\bf 117B}, 397 (1982).

\bibitem{frekan} J.~M.~Fr\`ere and G.~L.~Kane,
        Nucl. Phys. {\bf B223}, 331 (1983).

\bibitem{bartl89} A.~Bartl, H.~Fraas, W.~Majerotto, N.~Oshimo,
	Phys. Rev. {\bf D40}, 1594 (1989).

\bibitem{petcov} S.~T.~Petcov, Phys. Lett. {\bf 139B}, 421 (1984),
        S.~M.~Bilenky, N.~P.~Nedelcheva and S.~T.~Petcov,
	Nucl. Phys. {\bf B247}, 61 (1984);
        J.~Ellis, J.~M.~Fr\`ere, J.~S.~Hagelin, G.~L.~Kane
        and S.~T.~Petcov, Phys. Lett. {\bf 132B}, 436 (1983).

\bibitem{ref:fixpoint} M.~Carena and C.~E.~M.~Wagner,
        Preprint CERN-TH.7320/94 and references therein.

\bibitem{bartl92} A.~Bartl, H.~Fraas, W.~Majerotto, B.~M\"osslacher,
	Z. Phys. {\bf C55}, 257 (1992).

\bibitem{lahanas93} A.~B.~Lahanas, K.~Tamvakis and N.~D.~Tracas,
        Phys. Lett. {\bf 324B}, 387 (1994).

\bibitem{pierce93} D.~Pierce and A.~Papadopoulos,
        Phys. Rev. {\bf D50}, 565 (1994);
        Nucl. Phys. {\bf B430}, 278 (1994).

\bibitem{neutdec} E.~Reya, Phys. Lett. {\bf 133B}, 245 (1983);
	R.~Arnowitt, A.~H.~Chamseddine and P.~Nath,
        {\it ibid.} {\bf 129B}, 445 (1983);
	D.~Dicus, S.~Nandi, W.~Repko and X.~Tata,
        Phys. Rev. {\bf D29}, 1317 (1984); {\bf D30}, 1112 (1984);
	P.~Chiapetta, J.~Soffer, P.~Taxil, F.~M.~Renard and P.~Sorba,
	Nucl. Phys. {\bf B262}, 495 (1985);
	J.~Ellis, J.~S.~Hagelin, D.~V.~Nanopoulos and M.~Srednicki,
        Phys. Lett. {\bf 127B}, 233 (1983);
	V.~Barger, R.~W.~Robinett, W.~Y.~Keung and R.~J.~N.Phillips,
        {\it ibid.} {\bf 131B}, 372 (1983);
	G.~Altarelli, B.~Mele and S.~Petrarca,
        Nucl. Phys. {\bf B245}, 215 (1984);
	S.~Dawson, E.~Eichten and C.~Quigg,
        Phys. Rev. {\bf D31}, 1581 (1985).

\bibitem{raddec} H.~Komatsu and J.~Kubo,
        Phys. Lett. {\bf 157B}, 90 (1985);
        Nucl. Phys. {\bf B263}, 265 (1986);
        H.~E.~Haber, G.~L.~Kane and M.~Quir\'os,
        Phys. Lett. {\bf 160B}, 297 (1985);
        Nucl. Phys. {\bf B273}, 333 (1986);
        H.~E.~Haber and D.~Wyler,
        {\it ibid.} {\bf B323}, 267 (1989).

\bibitem{gunhab88} J.~F.~Gunion, H.~E.~Haber  \hbox{\it et al.}{},
        Int. J. Mod. Phys. {\bf A4}, 1145 (1987);
        J~.F.~Gunion and H.~E.~Haber, Phys. Rev. {\bf D37}, 2515 (1988).

\bibitem{bartl88-91} A.~Bartl, H.~Fraas, W.~Majerotto,
        Z. Phys. {\bf C41}, 475 (1988);
        A.~Bartl, W.~Majerotto, B.~M\"osslacher, N.~Oshimo,
        {\it ibid.} {\bf C52}, 477-485 and 677-684 (1991);
        A.~Bartl, W.~Majerotto, B.~M\"osslacher, N.~Oshimo, S.~Stippel,
        Phys. Rev. {\bf D43}, 2214 (1991).

\bibitem{pdg} ``Review of Particles Properties'', Particle Data Group,
        Phys. Rev. {\bf D50}, 1173 (1994).

\bibitem{ibanez} L.~E.~Ib\'a\~nez and C.~L\'opez,
        Nucl. Phys. {\bf B233}, 511 (1984);
	L.~E.~Ib\'a\~nez, C.~L\'opez and C.~Mu\~noz,
        {\it ibid.} {\bf B256}, 218 (1985);
        L.~E.~Ib\'a\~nez and G.~G.~Ross, Preprint CERN-TH.6412/92
        (1992), in ``Perspectives on Higgs Physics'',
        G.~L.~Kane (Ed.), p.229 and references therein.

\bibitem{deboer} W.~de Boer, Preprint IEKP-KA/94-01, hep-ph/9402266 (1994);
        W.~de Boer, R.~Ehret and D.~I.~Kazakov,
        Preprint  IEKP-KA/94-05, hep-ph/9405342 (1994);
        Phys. Lett. {\bf 334B}, 220 (1994).

\bibitem{barger} V.~Barger, M.~S.~Berger and P.~Ohmann,
	Phys. Rev. {\bf D47}, 1093 and 2038 (1993);
	V.~Barger, M.~S.~Berger, P.~Ohmann and R.~J.~N.~Phillips,
	Phys. Lett. {\bf 314B}, 351 (1993);
	V.~Barger, M.~S.~Berger and P.~Ohmann,
        Phys. Rev. {\bf D49}, 4908 (1994).

\bibitem{ref:higgs} H.~E.~Haber and R.~Hempfling,
        Phys. Rev. Lett. {\bf 66}, 1815 (1991);
	Y.~Okada, M.~Yamaguchi and T.~Yanagida,
        Prog. Theor. Phys. {\bf 85}, 1 (1991);
	Phys. Lett. {\bf 262B}, 54 (1991);
        J.~Ellis, G.~Ridolfi and F.~Zwirner,
	{\it ibid.} {\bf 257B}, 83 (1991);
        {\bf 262B}, 477 (1991);
        R.~Barbieri, M.~Frigeni and F.~Caravaglios,
        {\it ibid.} {\bf 258B}, 167 (1991);
        A.~Yamada, {\it ibid.} {\bf 263B}, 233 (1991).

\end{references}
\end{document}